\documentclass[aps,twocolumn,showpacs,amsmath,amssymb,floatfix]{revtex4}

\usepackage{dcolumn}
\usepackage{bm}

\usepackage{amsmath}
\usepackage{amsfonts}
\usepackage{amssymb}
\usepackage{array}
\usepackage[dvips]{graphicx}

\usepackage{color}

\usepackage{ulem}
\usepackage{xcolor}
\definecolor{DarkGreen}{rgb}{0,0.7,0}
\definecolor{DarkRed}{rgb}{0.5,0.0}
\definecolor{AB}{rgb}{0.6,0.0,0.6}
\definecolor{GS}{rgb}{0.8,0.0,0.0}

\newcommand{\Ci}{\mathrm{i}} 	
\newcommand{\e}{\mathrm{e}} 	

\newcommand{\bra}[1]{\langle #1 \vert}
\newcommand{\ket}[1]{\vert #1 \rangle}

\newcommand{\const}{\text{const}}

\newcommand{\op}{\hat}

\newcommand{\opH}{\op H}

\newcommand{\opcd}{\op c^\dagger}
\newcommand{\opcp}{\op c^{\phantom{\dagger}}}

\newcommand{\hc}{\text{H.c.}}

\renewcommand{\emph}[1]{\textit{#1}}

\renewcommand{\Im}{{\mathrm{Im}}}

\newcommand{\R}[1]{\mathrm{#1}}
\newcommand{\SD}{{\R{SD}}}
\newcommand{\VSD}{{V_\SD}}
\newcommand{\Vg}{{V_\R{g}}}

\newlength{\graphiclength}
\begin{document}

\preprint{Preprint}

\title{Conductance of correlated systems: real-time dynamics in finite systems}

\author{Alexander Bransch\"adel}
\affiliation{%
Institut f\"ur Theorie der Kondensierten Materie,  Karlsruher Institut f{\"u}r Technologie, 76128 Karlsruhe, Germany
}%
\author{Guenter Schneider} 
\affiliation{%
Department of Physics, Oregon State University, Corvallis, OR 97331, USA
}%
\author{Peter Schmitteckert}
\affiliation{Institut f\"ur Nanotechnologie, Karlsruher Institut f{\"u}r Technologie, 76344 Eggenstein-Leopoldshafen, Germany}

\date{\today}

\begin{abstract}
Numerical time evolution of transport states
using time dependent Density Matrix Renormalization Group (td-DMRG) methods has turned out to
be a powerful tool to calculate the linear and finite bias conductance of interacting impurity
systems coupled to non-interacting one-dimensional leads. Several models, 
including the Interacting Resonant Level Model (IRLM), the Single Impurity Anderson Model (SIAM), 
as well as models with different multi site structures, have been subject of investigations in this context. 
In this work we give an overview of the different numerical approaches that have been successfully applied to the problem and 
go into considerable detail when we comment on the techniques that have been used to obtain the full I--V-characteristics for the IRLM. 
Using a model of spinless fermions consisting of an extended interacting nanostructure 
attached to non-interacting leads, we explain the method we use to obtain the current--voltage characteristics and discuss the finite size
effects that have to be taken into account. We report results for the linear and finite bias
conductance through a seven site structure with weak and strong nearest-neighbor interactions.
Comparison with exact diagonalisation results in the non-interacting limit serve as a verification of the accuracy of our approach. 
Finally we discuss the possibility of effectively enlarging the finite system by applying damped boundaries
and give an estimate of the effective system size and accuracy that can be expected in this case.
\end{abstract}


\maketitle

\section{Overview}

	During the past decade improved experimental techniques have made the
	production of and measurements on one-dimensional systems possible
	\cite{Sohn_Kouwenhoven_Schon:1997}, and hence led to an increased
	theoretical interest in these systems. 
	However, the description of non-equilibrium transport properties,
	like the finite bias conductance of an interacting nanostructure 
	attached to leads, is a challenging task.
	In general, for non-interacting particles, the conductance can be extracted from the
	transmission of the  single particle levels \cite{Landauer57,Landauer70,Buettiker86}.
        For interacting particles in small or low-dimensional structures where the screening of 
	electrons is reduced, electron-electron correlations can no longer be neglected.
	Recently several methods to calculate the zero bias conductance of strongly interacting 
	nanostructures have been developed. One class of approaches consists in extracting the 
	conductance from an easier to calculate equilibrium quantity,
	e.g.\ the conductance can be extracted from a persistent current 
	calculation~\cite{Sushkov:PRB2001,Molina_EtAl:PRB2003,Meden_Schollwoeck:PRB2003,Molina_Schmitteckert_Weinmann_Jalabert_Ingold_Pichard:2004,Freyn_EtAl:X2009},
	from phase shifts in NRG calculations \cite{Oguri05},
	or from approximations based on the tunneling density of states \cite{MeirWingreenLee91}.
	Alternatively one can evaluate the Kubo formula within Monte-Carlo 
	simulations~\cite{Louis_Gros:2003}, 
	or from DMRG calculations~\cite{dan06,BohrSchmitteckert:PRB2007,Schmitteckert_Evers:PRL2008}.
   Linear conductance has also been investigated using Functional Renormalization Group 
	studies \cite{Karrasch_Enss_Meden:PRB2006}, or by diagonalizing small clusters and 
	attaching them to leads via a Dyson equation 
	\cite{Busser_Anda_Lima_Davidovich_Chiappe:PRB2000}.

	In contrast, there are only a few methods available to get rigorous results
	for the finite bias conductance. While the problem has been formally solved by
	Meir and Wingreen using Keldysh Greens functions~\cite{MeirWingreen92},
	the evaluation of these formulas for interacting systems is generally based on 
	approximations such as real time Keldysh RG~\cite{Schoeller_Koenig:PRL2000}.
	Within the framework of time dependent density functional theory (td-DFT) and Keldysh 
	Greens functions Stefanucci and 
	Almbladh~\cite{Stefanucci_Almbladh:PRB2004,Stefanucci_Almbladh:EPL2004}
	discuss the extraction of conduction from real time simulations.
	The restriction to finite sized systems for calculating transport within td-DFT was also 
	discussed by Di Ventra and Todorov~\cite{DiVentra_Todorov:JPCM2004}. 
	In \cite{Bushong_Sai_DiVentra:NL2005} Bushong, Sai, and Di Ventra discuss the extraction of 
	a finite bias current similar as discussed below in the framework of td-DFT.
	Weiss, Eckel, Thorwart and Egger \cite{PhysRevB.77.195316} discuss an iterative method 
	based on the summation of real-time path integrals (ISPI) in order to address quantum 
	transport problems out of equilibrium. Han and Heary \cite{Han_Heary:PRL2007} discuss 
	strongly correlated transport in the Kondo regime using imaginary time Quantum Monte Carlo techniques.
	
	In this work we review the concept of calculating the finite bias conductance of 
	nanostructures based on real time simulations 
	\cite{cazalilla02,luo03,daley04,white04,peter04,white05%
,SchneiderSchmitteckert06,Schmitteckert_HPC2007,Ulbricht_Schmitteckert_HPC2008,Branschaedel_Ulbricht_Schmitteckert:HPC2009%
,AlHassaniehFeiguinRieraBusserDagotto06,BoulatSaleurSchmitteckert2008,HeidrichMeisnerFeiguinDagotto09,HMeisnerMartinsBuesserAlHassanieh_etal2009%
,KirinoFujiiZhaoUeda08,SilvaHeidrichMeisnerFeiguinEtAl08}
	within the framework of the DMRG 
	\cite{White92,White93,Noack_Manmana:AIPCP2005,Hallberg:AIP2005,Schollwoeck:RMP2005}. %
	It provides a unified description of strong and weak interactions and works in the linear 
	and finite bias regime, 
	as long as finite size effects are treated properly. The method was successfully applied to 
	obtain results for the finite bias conductance in the 
	interacting resonant level model, showing perfect agreement with analytical methods based 
	on the Bethe ansatz \cite{BoulatSaleurSchmitteckert2008}. 
	I--V-characteristics have been obtained for the single-impurity Anderson model using the 
	adaptive td-DMRG-method \cite{HeidrichMeisnerFeiguinDagotto09}.
	Finite size effects and especially the impact of the possible combinations of tight 
	binding leads with 
	an even or odd number of sites coupled to the structure have been studied in detail in 
	\cite{HMeisnerMartinsBuesserAlHassanieh_etal2009} for a single impurity and for three 
	quantum dots. 
	Here, we show that finite size effects can be directly related to the structure of the 
	single particle energy levels in non-interacting systems.

	In a first approach of time dependent dynamics within DMRG, Cazalilla and Marston integrated
	the time-dependent Schr\"odinger equation in the Hilbert space
	obtained in a finite lattice ground state DMRG calculation~\cite{cazalilla02}.
	Since this approach does not include the density matrix for the time evolved
	states, its applicability is very limited.
	Luo, Xiang and Wang~\cite{luo03} improved the method by 
	extending the density matrix with the contributions of the wave function
	at intermediate time steps, restricting themselves to the infinite lattice algorithm.
	Schmitteckert~\cite{peter04} showed that the calculations 
	can be considerably improved by replacing the integration of
	the time dependent Schr{\"o}dinger equation with the evaluation of
	the time evolution operator using a  Krylov subspace method for matrix
	exponentials and by using the full finite lattice algorithm.
	
	\begin{figure}[t!]
		\includegraphics{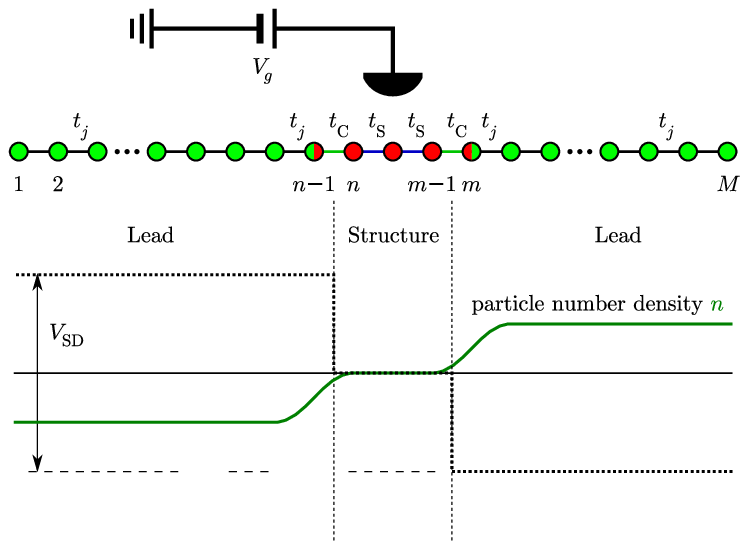}
		\caption{Interacting nanostructure \protect\includegraphics{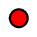} 
		attached to non-interacting leads \protect\includegraphics{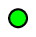} 
		(finite interaction $U_\R C$ with the first lead site 
		\protect\includegraphics{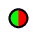} allowed) and schematic 
		density profile (green solid line) of the $N$-particle wavepacket at initial time $T=0$. 
		The density profile corresponds to the $N$-particle ground state of the Hamiltonian 
		$\opH+\opH_\R{SD}$, cf. Eq.~\eqref{eqn:Hamiltonian-SD}, where the bias voltage enters as 
		a local chemical potential $V_\R{SD}$ (black dotted line).}
		\label{Fig:NanoSystem}
	\end{figure}

	An alternative approach is based on wave function prediction
	\cite{WhitePredicition}. There one first calculates an initial state
	with a static DMRG. One iteratively evolves this state
	by combining the wave function prediction with a time evolution scheme.
	In contrast to the above mentioned full td-DMRG, one only keeps
	the wave functions for two time steps in each DMRG step.
	Different time evolution schemes have been implemented in the past using approximations 
	like the Trotter 
	decomposition~\cite{daley04,white04,AlHassaniehFeiguinRieraBusserDagotto06}, or the 
	Runge-Kutta method~\cite{white05}.
	Schneider and Schmitteckert~\cite{SchneiderSchmitteckert06,Schmitteckert_Schneider_HPC2006} 
	combined the idea of the adaptive DMRG method with
	direct evaluation of the time evolution operator via a matrix
	exponential using Krylov techniques as described in Ref.~\cite{peter04}. 
	Therefore the method involves no Trotter approximations,
	the time evolution is unitary by construction, and
	it can be applied to models beyond nearest-neighbor hopping.

	Concerning finite size effects, damped boundary conditions have been applied in order to 
	obtain an increased effective system size in the regime of small bias voltage 
	\cite{dan06,KirinoFujiiZhaoUeda08,SilvaHeidrichMeisnerFeiguinEtAl08},
	where an improved scheme for linear conductance was presented in 
	\cite{BohrSchmitteckert:PRB2007}.
	In the non-interacting case this can be traced back to a shift of the discrete single 
	particle energy levels of the system towards the center of the cosine band. 
	We demonstrate that this procedure can also be used when applying bias voltage of the order 
	of magnitude of the band width when handled carefully.

\section{The System}

	\begin{figure}[t]
		\includegraphics{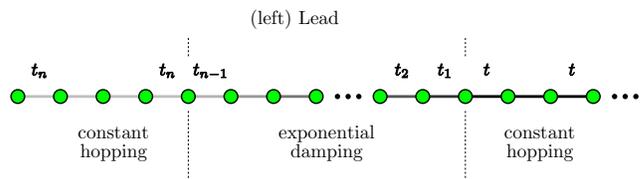}
		\caption{Exponential damping in the leads with $t_k=\Lambda^{-k/2}t$. In the damping 
		region, the hopping parameter is reduced by powers of the damping constant 
		$0<\Lambda^{-1/2}\leq 1$, while it is at the constant value $t$ where connected to the 
		nanostructure and at the constant value $\Lambda^{-n/2} t$ on the boundaries.}
		\label{Fig:damping}
	\end{figure}

	The Hamiltonian for the nanostructure is given by (S: the structure itself, L: leads, C: 
	contacts)
	\begin{equation}
		\label{eqn:Hamiltonian}
		\opH = \opH_\R{S} + \opH_\R{L} + \opH_\R{C},
	\end{equation}
	\begin{eqnarray}
		\label{eqn:Hamiltonian-system}
		\opH_{\R{S}} &=& 
		-\sum_{j=n+1}^{m-1} t_{\R{S}} ( \opcd_{j} \opcp_{j-1} + \hc )
			+ \sum_{j=n}^{m-1}V^{\phantom{g}}_{\R g j} \op n^{\phantom{g}}_j
		\nonumber\\ 
		& & 
		+\, \sum_{j=n+1}^{m-1} U \left(\op n_{j}-\frac{1}{2}\right)\left(\op n_{j-1}-\frac{1}{2}\right), \\
		\label{eqn:Hamiltonian-leads}
		\opH_{\R{L}} &=& 
			-\sum_{\substack{1 < j < n \\ m < j \leq M }} t_j ( \opcd_{j} \opcp_{j-1} +
											\hc), \\
		\label{eqn:Hamiltonian-contacts}
		\opH_{\R{C}} &=& 
		-t_{\R{C}} (\opcd_{n} \opcp_{n-1} + 
			\opcd_{m} \opcp_{m-1} + \hc) \nonumber \\
		& &
			+\, \sum_{j=n,m} U_\R C \left(\op n_{j}-\frac{1}{2}\right)\left(\op n_{j-1}-\frac{1}{2}\right),
	\end{eqnarray}
	where $\op n_j=\opcd_j\opcp_j$. 
	Individual sites are labeled according to Fig.~\ref{Fig:NanoSystem},
	$M_\R{Dot} = m - n$ is the size of the interacting nanostructure, $V_\R g$
	denotes a local external potential, which can be applied to the
	nanostructure, $U$ is a nearest-neighbor interaction inside the
	nanostructure, and $U_\R C$ is a nearest-neighbor interaction with the first lead sites. 
	The hopping elements in the leads, the structure, and
	coupling of the structure to the leads are $t_j$, $t_\R{S}$, and
	$t_\R{C}$, respectively. The hopping parameter in the leads $t_j$ is not necessarily 
	constant to allow for the inclusion of damped boundary-conditions. This can be used to divide the 
	leads in three areas, Fig.~\ref{Fig:damping}: here, two regions with constant hopping 
	matrix element $t$ and $\Lambda^{-n/2}t$ are smoothly coupled via a region of exponential 
	damped hopping, which allows for increasing the resolution of the level spacing of the 
	single particle energy levels on the energy scale $\Lambda^{-n/2}t$. For hard-wall 
	boundary-conditions, however, $t_j\equiv t= \const$.

	\begin{figure}[t!]
		\begin{center}
			\graphicspath{{.}}
			\includegraphics[width=0.95\linewidth]{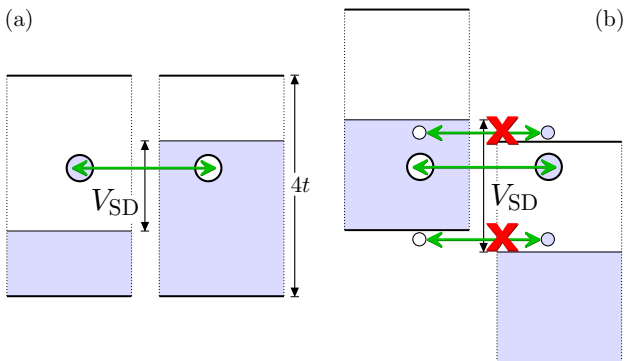}
			\caption{Different initial conditions, corresponding to (a) $\op H_\text{init.}=\op 
			H+\VSD(\op N_\text{L}-\op N_\text{R})/2$ and (b) $\op H_\text{init.}=\op H$. 
			The band width for the cosine band is $4t$. Assuming a single particle picture, we 
			understand that in case (a), increasing the bias voltage $\VSD$ to a value greater 
			than the band width qualitatively does not change the initial state, since all 
			particles populate only one of the two leads, while for case (b), quenching the leads 
			to different energies at the initial time prevents some particles (holes) from 
			tunneling from one lead to the other because of energy conservation. 
			For this reason there is no current flow in the extreme case of  $\VSD>4t$, cf. 
			Fig.~\ref{Fig:IV}.
			} 
		\label{fig:sketch_initial_conditions1}
		\end{center}
	\end{figure}

	The current operator $\op I_j$ at an arbitrary bond $j$ can be derived from the charge
	operator $\op Q_j=-e \op n_j$ using a continuity equation. For the tight-binding 
	Hamiltonian \eqref{eqn:Hamiltonian} the current operator and its expectation value take the 
	form
	\begin{eqnarray}
	\lefteqn{
		\op I_j 
		 = \Ci \frac e \hbar t_j \big [ 
			\opcd_j \opcp_{j+1} - \opcd_{j+1} \opcp_j
		  \big ] } \nonumber \\
	& \Rightarrow &
		I_j
		= -\frac{2e}\hbar t_j ~ \Im \bra{\Psi(T)} \opcd_j \opcp_{j+1} \ket{\Psi(T)}.
	\end{eqnarray} 
	We define the current
	through the nanostructure as an average over the current in the left and
	right contacts to the nanostructure
	\begin{equation} \label{eqn:current-average}
		I(T) = [ I_{n-1}(T) + I_{m-1}(T) ] / 2.
	\end{equation}

\section{Initial conditions and time evolution}
\label{ini_cond_plus_time_evo}
	\begin{figure}[t]
		\begin{center}
			\setlength{\graphiclength}{0.475\textwidth}
			\graphicspath{{./fig_data/data_local_FF_CURRENT/niRLM/}}
			\begin{footnotesize}\input{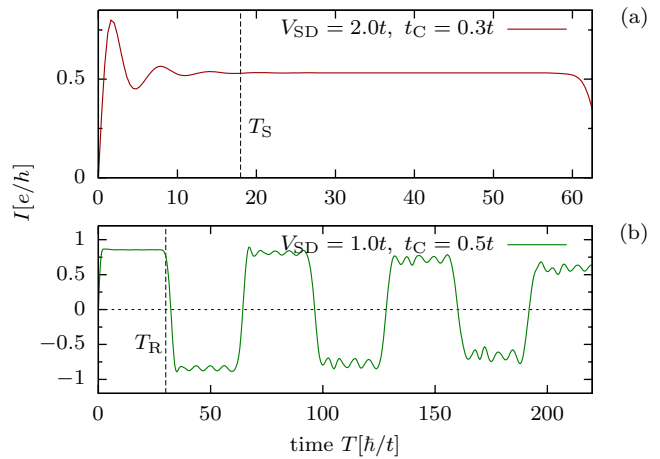}\end{footnotesize}
			\caption{Time dependent current through a single impurity coupled to noninteracting 
			1D leads for vanishing gate voltage $\Vg=0$. 
			The system consists of $M$ lattice sites and $N$ particles at nominal filling 
			$N/M=0.5$. We find three time domains: 1. an initial transient regime with decaying 
			oscillations, 2. a pseudo stationary current plateau and 3. finite size reflections. 
			(a) Shortly after the initial switching of the bias voltage the time dependent 
			behavior is dominated by oscillations which decay to a constant current plateau on 
			the time scale $T_\R S$ (here: $t_\R C=0.3t$, $M=120$).
			(b) The finite size of the system leads to reflections at the boundaries. A wave 
			packet that runs through the system starting at the impurity will be reflected at the 
			boundaries and returns to the impurity after time $T_\R R$. This results in the 
			typical pattern with recurrent sign changes of the current (here: $t_\R C=0.5t$, 
			$M=60$).}
			\label{Fig:TS_TR}
		\end{center}
	\end{figure}

	Following the prescription implemented in \cite{peter04,Ulbricht_Schmitteckert_HPC2008} we 
	add an external bias potential, namely the charge operator,
	\begin{equation}
		\opH_\R{SD} = \frac {\VSD} 2 \left( \sum_{j=1  }^{n-1}\,  \op n_j \; - \, \sum_{j=m}^{M} \op n_j \right)
		\label{eqn:Hamiltonian-SD}
	\end{equation}  
	to the unperturbed Hamiltonian $\opH$ and take the ground state 
	$\ket{\Psi_0}=\ket{\Psi(T=0)}$ of $\opH + \opH_\R{SD}$, obtained by a standard finite 
	lattice DMRG calculation, as initial state at time $T=0$ \cite{peter04}. The minimization 
	of the energy of the system leads to a charge imbalance in the right 
	(\underline{s}ource) and the left (\underline{d}rain) lead corresponding to $\VSD$, 
	as sketched in Fig.~\ref{fig:sketch_initial_conditions1}(a). 
	Alternatively, the bias voltage also can be added to the time evolution. 
	The initial state $\ket{\Psi_0}$ then has to be obtained as the ground state of the 
	unperturbed Hamiltonian $\opH$, 
	while the time evolution is performed using $\opH + \opH_\R{SD}$, cf. also 
	Fig.~\ref{fig:sketch_initial_conditions1}(b). 
	Starting from $\ket{\Psi_0}$, the time evolution of the system results from the time 
	evolution operator $\op U(T)$ 
	with $\ket{\Psi(T)}=\op U(T)\ket{\Psi_0}$,
	which leads to flow of the extended wave packet through the whole
	system until it is reflected at the hard wall boundaries as described
	in \cite{peter04}. 
	Corresponding to the two different schemes introduced before, $\op U$ is given as either 
	(a) $\op U(T)={\e}^{-\Ci \opH T/\hbar}$ or 
	(b) $\op U(T)={\e}^{-\Ci (\opH+\opH_\text{SD}) T/\hbar}$.

	The sudden switching of the bias voltage results in a ringing of the current in a
	transient time regime \cite{wingreen93}, see also Fig.~\ref{Fig:TS_TR}(a). 
	Here we show the short time behavior of the current through a single impurity coupled to 
	two leads in a system with $M=120$ lattice sites in total. This transient behavior with its
	characteristic oscillations decays on the time scale $T_\R S~\propto \Gamma$, where 
	$\Gamma$ is the width of the conductance peak. 
	By smearing out the voltage drop over a few lattice one may reduce the influence of large momentum states.
	Furthermore, the finite size of the system leads to 
	reflection of wave packets at the boundaries, cf. Fig.~\ref{Fig:TS_TR}(b). A wave packet
	travelling with Fermi velocity $v_\R F$ from the impurity towards the boundaries will return
	to the impurity after a transit time given by $T_\R R \propto M/v_\R F$, which is the 
	characteristic time scale for finite size effects appearing in the expectation value of time 
	dependent observables.

	\begin{figure}[t]
		\begin{center}
			\setlength{\graphiclength}{0.475\textwidth}
			\graphicspath{{./fig_data/data_local_FF_CURRENT/niRLM/}}
			\begin{footnotesize}\input{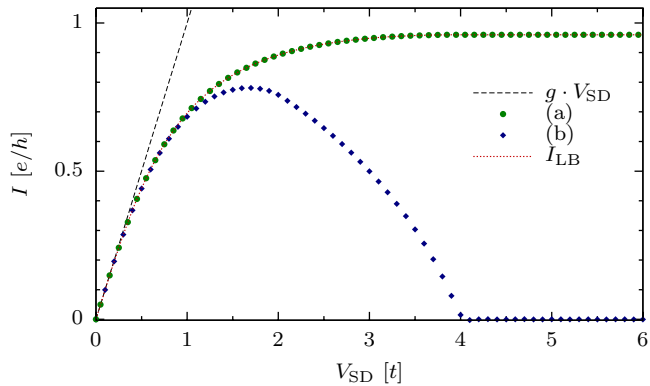}\end{footnotesize}
			\caption{
				I--V-characteristics for the resonant level model with 
				$t_\text{C}=0.4t$ and $U_\text{C}=0$. 
				The linear conductance is $1$. The plot shows results for two different time 
				evolution schemes: 
				(a) the initial state $\ket{\Psi_0}$ of the system is the ground state of the 
				    Hamiltonian $\opH+\VSD(\op N_\text{L}-\op N_\text{R})/2$, 
				    while the time evolution is performed as 
					 $\ket{\Psi(T)}=\exp(-\Ci\opH T)\ket{\Psi_0}$. 
				(b) the initial state $\ket{\Psi_0}$ of the system is the ground state of the 
				    Hamiltonian $\opH$, 
				    while the time evolution is performed as 
				    $\ket{\Psi(T)}=\exp[-\Ci(\opH+\VSD(\op N_\text{L}-\op N_\text{R})/2)T]\ket{\Psi_0}$. 
				For values of the bias voltage much smaller than the band width the both 
				approaches agree nicely. 
				However, we find strong deviations when band edge effects come into play. 
				Note that (a) corresponds to the situation of wide band metallic leads.
				Since our emphasis  lies on the description of nanostructures attached to metallic 
				leads we prefer to work in this approach.
				When describing situations with band gap materials as leads one should refer to
				approach (b).
				For further discussion see Fig.~\ref{fig:sketch_initial_conditions1} and the text.
			}
		\label{Fig:IV}
		\end{center}
	\end{figure}

	To compare the approaches (a) and (b), we show  current voltage-characteristics in 
	Fig.~\ref{Fig:IV} for the resonant level model with a single 
	impurity ($M_\text{Dot}=m-n=1$, cf. Fig.~\ref{Fig:NanoSystem}) coupled to two leads via the 
	hopping matrix element $t_\R C=0.4t$ and the gate voltage as well as the interaction set to 
	$U_\text{C}=\Vg=0$. The dots correspond to results obtained numerically using exact diagonalisation, 
	while the lines 
	correspond to analytic calculations included for comparison. Here, the straight line shows the
	current assuming linear scaling with $\VSD$ with linear conductance $g=1$, while the curved line
	overlaid by the numerical results for approach (a) has been obtained using the Landauer--B\"uttiker
	approach, taking cosine-dispersion into account.

	\begin{figure}[t]
		\begin{center}
			\setlength{\graphiclength}{0.475\textwidth}
			\graphicspath{{./fig_data/data_local_DMRG_CURRENT/}}
 			\begin{footnotesize}\input{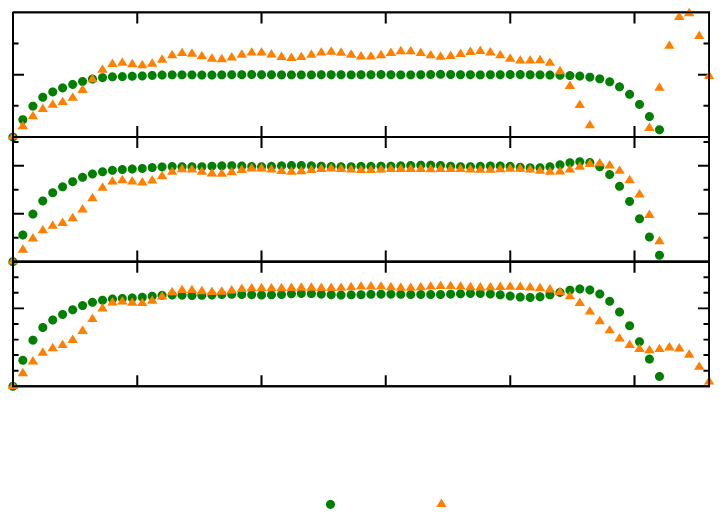}\end{footnotesize}
			\caption{Time dependent current through a single impurity coupled to noninteracting 
			1D leads with $t_\R C=0.4t$ and $U_\R C=2.0t$ 
			for different values of $\VSD$ and vanishing gate voltage $\Vg=0$. 
			The system consists of $M=48$ lattice sites and $N$ particles at nominal filling 
			$N/M=0.5$. 
			The current is obtained from a td-DMRG calculation by performing the time evolution 
			on an initial non equilibrium state, using a DMRG projection scheme with a variable 
			number of kept states $100\leq N_\mathrm{cut}\leq 5600$ with the discarded entropy 
			$S_\mathrm d$ kept below a certain value (here: $S_\mathrm d\lesssim 10^{-3}$; cf. 
			also Fig.~\ref{Fig:compare_NCUT_connected_disconnected}).
			(a) The initial state $\ket{\Psi_0}$ corresponds to the situation sketched in 
				Fig.~\ref{fig:sketch_initial_conditions1}(a) 
			   where $\ket{\Psi_0}$ is obtained as the ground state of 
				$\op H_\text{init.}=\op H+\VSD(\op N_\text{L}-\op N_\text{R})/2$, 
			(c) The initial state $\ket{\Psi_0}$ is obtained as the ground state of 
				$\op H_\text{init.}\big\vert_{t_\R C=0, U_\R C=0}$. 
				The current plateau we are looking for can be obtained more reliable when using 
				prescription (a).
			}
		\label{Fig:compare_connected_disconnected}
		\end{center}
	\end{figure}

	The procedure of 
	extracting the current from the numerical data will be described in the next section. Here 
	we want to emphasize the different results we get for the I--V-curve for the two different 
	cases. 
	For the tight binding Hamiltonian the dispersion relation is given by
	$\epsilon_k=-2t \cos k$, with a finite band width $4t$. 
	For the approach (a) this leads in the non-interacting case  to a saturation of $I(\VSD)$ 
	for all values of the bias voltage $\VSD\ge 4t$.
	Further increasing $\VSD$ beyond the band edge does not  change the initial occupation of 
	energy levels.
	In contrast, for the case (b), the particles will be distributed equally over the left and 
	the right lead in the initial state $\ket{\Psi_0}$, whereas the voltage enters in the time 
	evolution operator. 
	For small values of $\VSD$ we find a good agreement for $I(\VSD)$ for (a) and (b), while 
	for $\VSD \gtrsim 2t$ there is a mismatch which finds its expression in a current maximum 
	for $0<\VSD<4t$ with a subsequent break down to $I=0$ for $\VSD > 4t$. 
	This behavior has been predicted in \cite{Cini80} and can be understood from
	Fig.~\ref{fig:sketch_initial_conditions1}(b), which explains how energy 
	conservation prevents particles (holes) to tunnel from one lead to the other which removes 
	contributions to the current. \footnote{We want to emphasize that the negative differential 
	conductance for 
	the IRLM with tight binding chains in \cite{BoulatSaleurSchmitteckert2008} is not 
	related to the band effect described here. In fact, approach (a) has been used there 
	for the numeric simulation while, in contrast, we find saturation of the current 
	in the non-interacting case. In addition the maximum of the current appears at an energy 
	below half the band width, where both approaches give the same result.}. 
	More recently, a detailed analysis of the negative differential
	conductance for the situation (b) has been carried out \cite{BaldeaKoppel2010}. In this work, it has been 
	realised that the density of states in the leads adds a major contribution to the breakdown
	of the current.

	\begin{figure}[t]
		\begin{center}
			\setlength{\graphiclength}{0.475\textwidth}
			\graphicspath{{./fig_data/data_local_DMRG_CURRENT/}}
 			\begin{footnotesize}\input{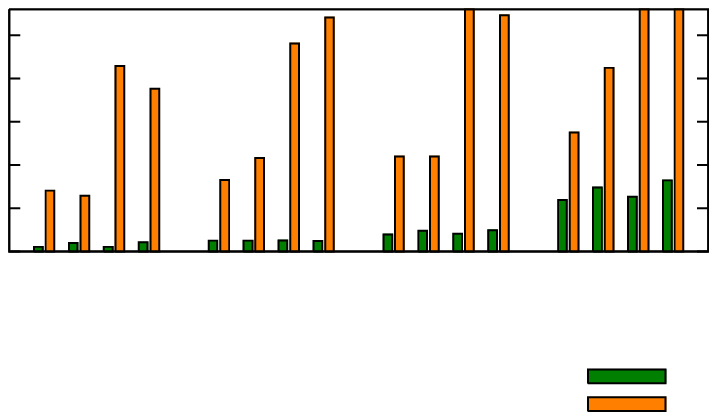}\end{footnotesize}
			\caption{
			Maximum dimension $N_\mathrm{cut}$ of the DMRG projection scheme for an 
			I--V-calculation necessary to keep the discarded entropy $S_\text{d}$ 
			below a certain value (here: $S_\text{d}\lesssim 10^{-3}$) for different 
			configurations I to IV and different values of the bias voltage $\VSD$,
			where we used $100\leq N_\mathrm{cut} \leq 5600$ states as a second 
			limitation. Here, the current through the contact links to a single  
			impurity with $t_\mathrm{C}=0.4t$ is obtained for 70 time steps 
			($\Delta T=0.4\hbar/t$) in a system with $M=48$ lattices sites at half 
			filling. 
			(a) The initial state $\ket{\Psi_0}$ is the ground state of 
			$\op H_\text{init.}=\op H+\VSD(\op N_\text{L}-\op N_\text{R})/2$, 
			(c) $\ket{\Psi_0}$ is obtained as the ground state of 
			$\op H_\text{init.}\big\vert_{t_\R C=0, U_\R C=0}$.
			}
		\label{Fig:compare_NCUT_connected_disconnected}
		\end{center}
	\end{figure}

	Moreover, there are other approaches to how the initial state and the time evolution can be 
	defined. For example, in addition to prescription (a), the coupling $t_\R C$ and the 
	interaction $U_\R C$ can be set to zero for the calculation of $\ket{\Psi_0}$. In this 
	case (c), 
        both leads as well as the structure are totally independent systems, and there is a 
	very intuitive connection of $\VSD$ and the difference of the particle number in the left 
	and the right lead, because the isolated leads can be described in a single particle 
	picture. The drawback of this approach, which adds a sudden switching of $t_\R C$ and 
	$U_\R C$ in addition to the switching of $\VSD$ at initial time $T=0$, is an enhanced 
	transient regime and therefore a reduced plateau of constant current that we need to 
	extract the I--V-curve from. In Fig.~\ref{Fig:compare_connected_disconnected} we compare
	the time dependent current obtained using the different initial conditions (a) and (c) for
	a single impurity coupled to two leads via $t_\R C=0.4t$, including a finite density-density
	interaction $U_\R C=2.0t$, for different values of $\VSD$. To evaluate the time evolution
	of a system with finite interaction numerically, we used the td-DMRG method, with parameters
	as described in the figure caption of Fig.~\ref{Fig:compare_connected_disconnected}. For 
	both approaches (a) and (c), we find a time regime of (quasi) constant current. 
        However, approach (a) has several advantages over (c): the current plateau is more consistent, 
        which simplifies analysis, and to keep the discarded entropy $S_\text{d}$ in the td-DMRG 
        calculation below a predefined threshold, the number of states, which have to be kept in the DMRG, 
        is considerably higher for (c) when compared to (a), making approach (c) computationally much more 
        expensive. The latter point is illustrated in Fig.~\ref{Fig:compare_NCUT_connected_disconnected}, 
	where we compare the maximum dimension $N_\R{cut}$
	of the DMRG projection scheme that is necessary to keep $S_\text{d}\lesssim 10^{-3}$, for different
	values of the bias voltage $\VSD$, of the gate voltage $\Vg$ and of the interaction $U_\R C$. 
	We always find a much smaller value of $N_\R{cut}$ for (a) as compared to (c).

	Another problem of approach (c) is the discretization of the I--V-curve into steps resulting from the 
	discrete single particle energy levels of the initial state. This could probably be handled 
	using a procedure similar to the one described in section \ref{Sec:DensityShift}.

	For these reasons we will use approach (a) throughout the remainder of this paper.

\section{Differential and linear conductance} 

	\begin{figure}[t]
		\begin{center}
			\setlength{\graphiclength}{0.475\textwidth}
			\graphicspath{{./fig_data/data_local_FF_CURRENT/niRLM/}}
			\begin{footnotesize}\input{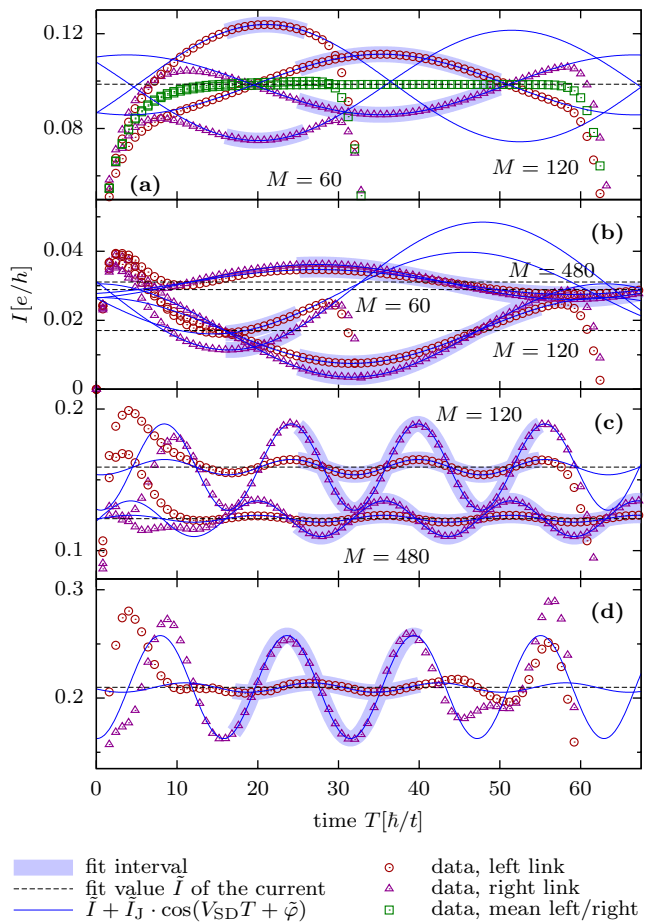}\end{footnotesize}
			\caption{Current through a single impurity with $t_\R C=0.3t$ at nominal filling 
			$N/M=0.5$ obtained from exact numerical diagonalization (a-c), or DMRG including 
			interaction (d), respectively -- (a) for different system sizes $M$ at bias voltage 
			$V_\R{SD}=0.1t$ and gate voltage $\Vg=0$. The black dashed line corresponds to the 
			mean value of the fit values $\tilde I$ for the left 
			and right contact link, for $M=60$ lattice sites. The fit interval has 
			to be chosen carefully -- initial oscillations from the bias voltage switching and 
			the finite transit time have to be taken into account. Even though the period of the finite size 
			oscillations considerably exceed the system size $M=60$ for $V_\R{SD}=0.1$, the fit 
			current $\tilde I$ is in nice agreement with the current plateau of the $M=120$ 
			system. However, finite size effects still have to be addressed (b, $\Vg=0.3t$, 
			$V_\R{SD}=0.1t$, and c, $\Vg=0.3t$, $V_\R{SD}=0.4t$) 
			since in general the fit current can strongly depend on the system size -- in particular, a 
			non-zero gate voltage changes the particle number density in the leads when the 
			overall particle number is fixed. The same fit procedure can be applied to 
			interacting systems (d, $U_\R C=2.0t$, $V_\R{SD}=0.4t$, $\Vg=0.3t$).}
		\label{Fig:RLM_fit}
		\end{center}
	\end{figure}
\begin{figure}[t]
\begin{center}
  \setlength{\graphiclength}{0.475\textwidth}
  \graphicspath{{./fig_data/data_local_FF_CURRENT/I_J_M/}}
  \begin{footnotesize}\input{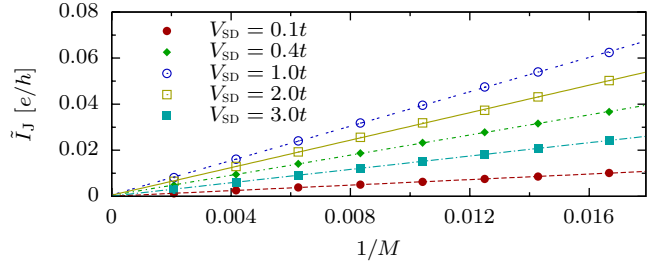}\end{footnotesize}
  \caption{Oscillation amplitude $\tilde I_\R J$ from fits as shown in Fig.~\ref{Fig:RLM_fit}, as 
  function of the inverse system size $1/M$ for different values of $\VSD$, of the time
  dependent current through a single contact link to a single impurity, with $t_\R C=0.5t$
  and $\Vg=0$.
  }
  \label{Fig:I_J_M}
\end{center}
\end{figure}

	For the calculation of the DC-conductance through the
	nanostructure the 
	time evolution has to be carried out for sufficiently long times
	until a quasi-stationary state is reached and the steady state
	current $I$ can be calculated.
	If the stationary state corresponds to a well-defined applied external
	potential $\VSD$, the differential conductance is given by
	$g(\VSD) = e\, \partial I(\VSD) / \partial \VSD.$ 
	In the limit of a small applied potential, $\VSD
	\rightarrow 0$, the linear conductance is given by 
	$g(\VSD) = e I(\VSD) / \VSD.$

\begin{figure}
\begin{center}
  \setlength{\graphiclength}{0.475\textwidth}
  \graphicspath{{./fig_data/data_local_FF_CURRENT/niRLM/}}
  \begin{footnotesize}\input{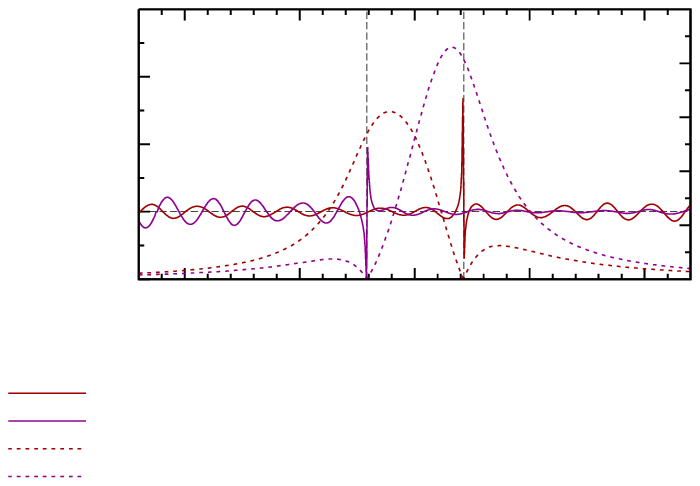}\end{footnotesize}
  \caption{Fit of the oscillation frequency $\tilde\omega_\R J=2\pi/\tilde T_\R J$ of the 
  Josephson oscillations in a system with $M=120$ lattice sites and a single resonant level 
  with $t_\R C=0.3t$ at a bias voltage $\VSD=0.4t$. 
  The oscillation period extracted from the time evolution of the current is in excellent 
  agreement with the analytical expression $\omega_\R J=\vert\VSD\vert/\hbar$ (dashed black 
  line). 
  The kinks that appear in $\tilde \omega_\R J$ can be traced back to the fact that the 
  amplitude of the oscillations $\tilde I_\R J$ vanishes for $\Vg\approx\pm\VSD/2$ at either 
  the left or the right contact link. 
  Then a fit of $\tilde \omega_\R J$ does not work. The residual wiggling (its amplitude as well 
  as its frequency) depends on the size and the position of the fit interval 
  $[T_\text{min}, T_\text{max}]$, 
  and is therefore consistent with a finite fitting interval in time domain. Enlarging the fit 
  intervall in conjunction with the system size reduces this effect (not shown here).
  }
  \label{Fig:VSD_fit}
\end{center}
\end{figure}

	To discuss the general behavior of the time evolution from an initial nonequilibrium state 
	we first consider the most simple case we can think of: transport through a single impurity.
	The current rises from zero and settles into a quasi-stationary
	state, Fig.~\ref{Fig:TS_TR}(a). After the wavepackets have traveled to the boundaries of the
	system and back to the nanostructure, the current falls back to zero
	and changes sign, cf. Fig.~\ref{Fig:TS_TR}(b). Additionally there is a third type of finite
	size oscillations, Fig.~\ref{Fig:RLM_fit}. Here we show the time dependent current for different
	configurations, from the leads to the impurity on a single (left or right) contact link, and through 
	the impurity as defined in Eq.~\eqref{eqn:current-average}.
	After the initial oscillations have decayed on the time scale $T_\R S$, the current through a single
	contact link shows remaining oscillations, with an amplitude depending on $\VSD$ and $\Vg$,
	and proportional to the inverse of the system size $1/M$. The latter is demonstrated in Fig.~\ref{Fig:I_J_M}.
	The period of the oscillation  depends on the applied bias voltage
	[compare Fig.~\ref{Fig:RLM_fit} (b, c)] but is independent of
	the system size [Fig.~\ref{Fig:RLM_fit} (b-d)] and of the gate potential [Fig.~\ref{Fig:VSD_fit}],
	and is given by $T_{\mathrm{J}}=2\pi\hbar/\vert \VSD \vert$. 
	In the resonant tunneling case [Fig.~\ref{Fig:RLM_fit}(a), $\Vg=0$], the oscillations on the left
	and the right contact link cancel in the current average Eq.~\eqref{eqn:current-average} due to a 
	different sign in the amplitude of the oscillations $\tilde I_\R J$, which
	does not hold in general [Fig.~\ref{Fig:RLM_fit}(b-d), $\Vg\neq0$], where the amplitude of the 
	oscillations as a function of the gate potential $\Vg$ varies differently on the individual contact 
	links, Fig.~\ref{Fig:VSD_fit}.

	In Fig.~\ref{Fig:VSD_fit} we plot the fit of the oscillation frequency 
	$\tilde \omega_\R J=2\pi/\tilde T_\R J$ as a function of the gate potential $\Vg$ for a 
	fixed value of $\VSD$, where we find $\tilde \omega_\R J$ to be independent of the gate 
	potential. To be precise, the fit nicely confirms the above relation of $\VSD$ and oscillation period.
	This periodic contribution to the current is reminiscent of the Josephson 
	contribution in the tunneling Hamiltonian, obtained by
	gauge transforming the voltage into a time dependent coupling
	$\tilde{t}_{\mathrm{C}}(T) = {t}_{\mathrm{C}} \,{\mathrm e}^{\R i \VSD T/\hbar}$ 
	\cite{Mahan:MPP2000}.
	Like in a tunnel barrier in a superconductor, we have a phase coherent quantum system,
	namely the ground state at zero temperature. Instead of the superconducting gap we have a
	finite size gap resulting from the finite nature of the leads. Therefore the amplitude of this 
	residual wiggling vanishes proportional to the finite size gap provided by the leads.
	
	The stationary current is given by a fit to 
	$\tilde I+ \tilde{I}_\R J \cos(2\pi T/T_{\mathrm{J}}+\tilde\varphi)$ 
	with the fit-parameters tagged by a tilde, 
	because the oscillation period $T_{\mathrm{J}}$ is known.
	In general, the density in the leads, and therefore also the
	current, depends on the system size and a finite size
	analysis has to be carried out in order to extract quantitative results
	[Fig.~\ref{Fig:RLM_fit} (b,c), see also discussion of
	Fig.~\ref{Fig:gnf-d7v0t80t50}].
	Only in special cases (symmetry, half filled leads, and zero gate
	potential) is the stationary current independent of the system size
	[Fig.~\ref{Fig:RLM_fit} (a)].

\section{Finite size effects}
\label{sec:finite_size_effects}

  Finite size effects such as the finite transit time of a wave packet traveling through the 
  system and the periodic contribution to the current make it difficult to obtain a 
  pseudo-stationary state where a constant current can be extracted from the time evolution. 
  This problem can be treated by a fit procedure as discussed in the previous section. However, in the small 
  bias regime, where the amplitude of the oscillations is bigger than the (expected) current 
  and the oscillation time $T_\R{J}$ exceeds the transit time, this approach is unreliable. 
  In section \ref{Sec:ExpDamp} we discuss the possibility of effectively enlarging the system 
  using damped boundary conditions (DBC) while keeping the system size $M$ constant (cf. 
  Fig.~\ref{Fig:damping}). Furthermore, the time evolution of the current strongly depends on 
  the number of lattice sites of the leads being \underline{e}ven or \underline{o}dd, 
  Figs.~\ref{Fig:twoimpurities_compare_size}, \ref{Fig:twoimpurities_compare_even_odd}. In 
  Fig.~\ref{Fig:twoimpurities_compare_size} we compare this effect for 
  a non-interacting two-dot structure for different system sizes in the regime of very small 
  voltage $\VSD\ll t$, where we consider three qualitatively different cases, 
  (a) $T_\R R\ll T_\R J$, (b) $T_\R R \approx T_\R J$ and (c) $T_\R R\gg T_\R J$, where $T_\R R$,$T_\R J$ 
  denote the transit time and oscillation period respectively, as discussed in Sec.\ref{ini_cond_plus_time_evo}. 
  Since the
  number of single particle energy levels is equal to the number of lattice sites,
  these relations are connected to $\VSD$ and the level spacing $\Delta \epsilon$ as,
  (a) $\Delta\epsilon\gg\VSD$, (b) $\Delta\epsilon\approx\VSD$ and (c)
  $\Delta\epsilon\ll\VSD$. Intuitively one would expect that the level discretisation
  must be small compared to the energy scales of interest, and indeed we find, that on the 
  time scale $T<T_\R R$ the numerical simulation fits best with the analytic result $I_\R{LB}$ 
  obtained from the Landauer--B\"uttiker approach in case (c) (see Fig.\ref{Fig:twoimpurities_compare_size}). 
  However, in all cases, the time
  evolution of the current depends on the different configurations of the leads with even
  or odd number of lattice sites. Two aspects must be distinguished: (1) the 
  qualitative difference in the time evolution depending on wether the number of lead sites is 
  equal (as for the e2e and the o2o configuration), or unequal (as for the e2o and the o2e
  configuration), is clearly demonstrated in the figure. For the 
  two-dot structure, this holds true even for $T_\R R\gg T_\R{J}$, 
  Fig.~\ref{Fig:twoimpurities_compare_size}~(c). For the o2o and the e2e configurations we 
  find a behavior where the current suddenly increases by a factor of $\sim2$ after the 
  transit time $T_\R R$, as opposed to the ``expected'' behavior with a sign change, seen for the o2e and the e2o 
  configuration. (2) An overall odd number of lattice sites $M$ (e.g. the o2e
  and the e2o configurations) shifts the filling factor in the leads away from $0.5$ due to 
  their finite size. A similar effect results from applying a gate voltage $\Vg\neq0$, which 
  imposes a problem to the extraction of the linear conductance. A possible solution is discussed
  in Sec.~\ref{Sec:DensityShift}.

%

\subsection{Even-odd effect}

  In \cite{HMeisnerMartinsBuesserAlHassanieh_etal2009}, a detailed analysis 
  of finite size effects resulting from an even or odd number of lattice sites 
  in the leads for a single-dot and for a three-dot structure with on-site 
  interaction including the spin degree of freedom has been carried out. 
  The behavior of the time dependence of the current resulting from the type of 
  the lead (\underline{e}ven or \underline{o}dd number of sites) 
  has been traced back to the different magnetic moment of the system which 
  is $S^z_\R{total}=1/2$ for an overall odd number $M$ of lattice sites 
  and $S^z_\R{total}=0$ for $M$ being even. The reduction of the current in a 
  situation where the leads both consist of an even number of sites (e$n$e) 
  as compared to the other possible combinations (o$n$e, o$n$o) has been 
  explained by the accumulation of spin on the structure in the first case 
  corresponding to the effect of applying an external magnetic field.

\begin{figure}[tb]
\begin{center}
  \setlength{\graphiclength}{0.475\textwidth}
  \graphicspath{{./fig_data/data_local_FF_CURRENT/twoimpurities/}}
  \begin{footnotesize}\input{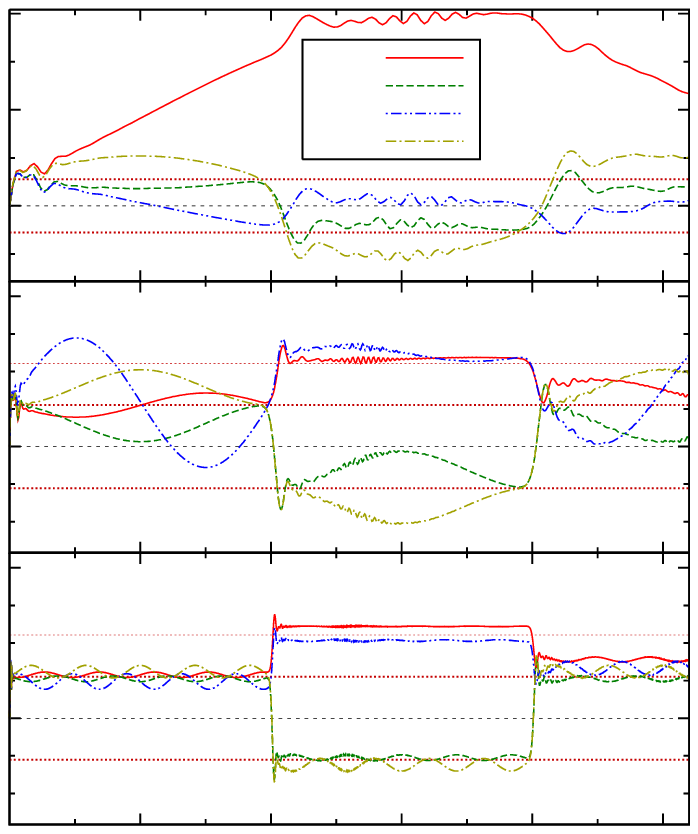}\end{footnotesize}
  \caption{
  Current through the contact link of a structure with two dots ($t_\R S=t$), coupled to leads 
  with a finite number of sites $M$ and $t_{\mathrm{C}}=0.5t$ (compare also Fig.~\ref{Fig:NanoSystem}), at 
  nominal half filling $N/M = 0.5$ obtained from exact numerical diagonalization for bias 
  voltage $\VSD=0.05 t$. The horizontal dotted lines represent the  analytical result 
  $I_\text{LB}$ obtained from the Landauer--B\"uttiker approach. The current is measured on 
  the left link to the structure. The time axis is normalized to the transit time 
  $T_\R R = M\hbar/(2t)$. Here, the focus is on finite size effects in the low voltage 
  regime. We distinguish three cases: the system size is very small in case 
  (a) where $M=60+x$ with $x=0$ ($29$ lattice sites on the left and right which is an 
  \underline{o}dd number in both cases o2o), $x=1$ (now $30$ sites on the left which is an 
  \underline{e}ven number e2o), $x=2$ (e2e) and $x=3$ (o2e). Here, the single particle level 
  spacing $\Delta \epsilon$ is much longer than $\VSD$, while the period of the Josephson 
  oscillations $T_\R J=2\pi\hbar/\vert\VSD\vert$ is much bigger than the transit time 
  $T_\R R$. 
  Case (b) shows an intermediate situation with $M=252+x$ lattice sites. Here, 
  $\Delta \epsilon\approx\VSD$ and $T_\R J \approx T_\R R$. A situation where 
  $\Delta \epsilon<\VSD$ and $T_\R J < T_\R R$ is realized in case (c) with $M=1200+x$. For 
  the e2o and the o2e case one has to do a density shift correction of the result since the 
  total number of particles $N\neq M/2$, cf. Sec. \ref{Sec:DensityShift}.
  }
  \label{Fig:twoimpurities_compare_size}
\end{center}
\end{figure}

\begin{figure}[tb]
\begin{center}
  \setlength{\graphiclength}{0.475\textwidth}
  \graphicspath{{./fig_data/data_local_FF_CURRENT/}}
  \begin{footnotesize}\input{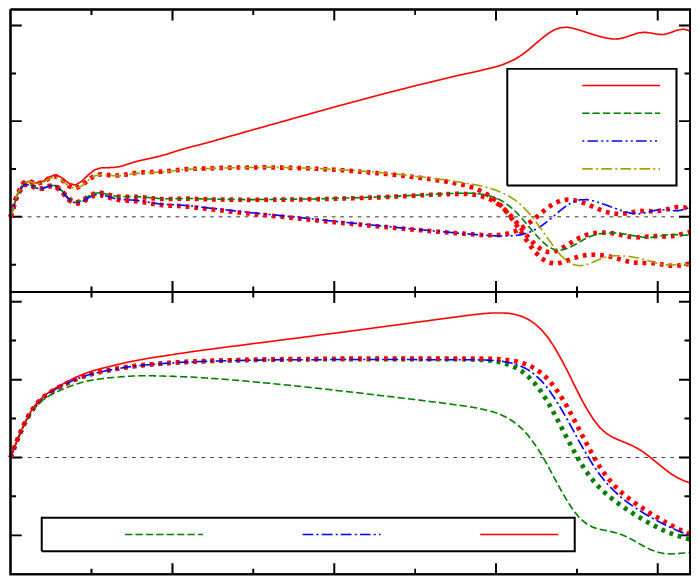}\end{footnotesize}
  \caption{
  Current through a structure coupled to two leads (mean value of left and right contact link) 
  with an overall finite system size $M$ at half filling obtained from exact diagonalization. 
  The figure demonstrates the influence of the number of lattice sites in the leads 
  (\underline{e}ven or \underline{o}dd) on the current for a bias voltage $\VSD$ smaller than 
  the single particle level spacing. 
  The dotted lines represent a situation where an additional constant voltage $\Delta V$ is 
  applied to both leads (a) or to the left lead (b), respectively. 
  $\Delta V \neq 0$ results in a shift of the single particle levels in the uncoupled leads 
  which can be used to ``mimic'' the different combinations of leads with an even or odd number 
  of lattice sites. 
  (a) $M=60+x$, $x=0$ (o2o), $1$ (e2o), $2$ (e2e) and $3$ (o2e) where the number of electrons 
  is $N=30$ for $M=60,61$ and $N=31$ for $M=62,63$. The dotted lines all together are generated 
  using a system with $M=60$ lattice sites, with $\Delta V\neq 0$. The different situations e2o 
  and o2e can be recovered by changing the particle number from $N=30$ to $N=31$, cf.
  Sec.~\ref{Sec:DensityShift}. 
  (b) $M=61+x$, $x=0$ (e1e), $1$ (o1e) and $2$ (o1o) where the particle number is fixed to 
  $N=31$. Here, the green (red) dotted line is generated from the e1e (o1o) system.
  }
  \label{Fig:twoimpurities_compare_even_odd}
\end{center}
\end{figure}

  We already find parity effects in the time dependence of noninteracting spinless fermions
  in a system with a single-dot or a two-dot structure, Figs.~\ref{Fig:twoimpurities_compare_size},
  \ref{Fig:twoimpurities_compare_even_odd}.
  In the following we will trace the parity effects back to the level structure in the leads.
  The single particle levels $\epsilon_k$ of an uncoupled, noninteracting lead with $M_i$ sites ($i=\R L, \R R$) 
  are given by $\epsilon_k=-2t\cos [\pi k/( {M_i+1} )]$,  
  $k=1,\ldots,M_i$.
  The energy of a particle residing on a decoupled single dot structure ($t_\R C =0$) 
  is simply given by the gate voltage $\epsilon_\R d=V_\R g$, which is at the Fermi edge for $V_\R g=0$. 
  For a decoupled $n$-dot structure one gets $\epsilon_{\R d,j}=-2t_\R S \cos [\pi j/(n+1)]+V_\R g$, $j=1,\ldots,n$. 
  For an equal number of sites on both leads (as for example e$n$e or o$n$o) 
  there is a twofold degeneracy of the single particle lead levels 
  which does not exist if $M_\R L = M_\R R \pm 1$. In the degenerate case, 
  single particle eigenfunctions can be constructed with a fully
  delocalized particle density while for $M_\R L = M_\R R \pm 1$, the density 
  profile of the single particle wave functions shows an alternating 
  confinement of the particle on either the left or the right lead 
  The same 
  holds true for the energy levels of the structure: 
  if degenerate with a lead level, the single particle wave function can 
  be distributed over the whole lead while it is localized on the structure otherwise. 
  Therefore, in the e1e case, the single-dot level is not degenerate with the 
  lead levels when $\epsilon_\R d=0$. 
  As a result, a single particle occupying the dot level generates a sharp 
  peak in the density profile (as well as the spin profile). 
  For the o1o case on the other hand, both leads have one energy level in 
  the middle of the band, which together with the dot level generates a threefold degeneracy. 
  For finite coupling $t_\R C>0$, the degeneracy of the lead levels and of 
  the levels of the structure with the lead levels gets lifted. 
  The single particle wave functions must be divided equally on both leads, 
  when $M_\R L=M_\R R$, while the alternating confinement is preserved 
  for $M_\R L = M_\R R \pm 1$. Concerning the energy level of the dot, 
  the threefold degeneracy in the uncoupled o1o case results in two levels 
  with strong localization on the dot, one lifted above the Fermi edge and 
  one pushed below, and a third level with vanishing particle density on the dot, 
  remaining on the Fermi edge. 

\begin{figure}[t!]
\begin{center}
  \setlength{\graphiclength}{0.475\textwidth}
  \graphicspath{{./fig_data/data_local_FF_CURRENT/Eigenvalues/}}
  \begin{footnotesize}\input{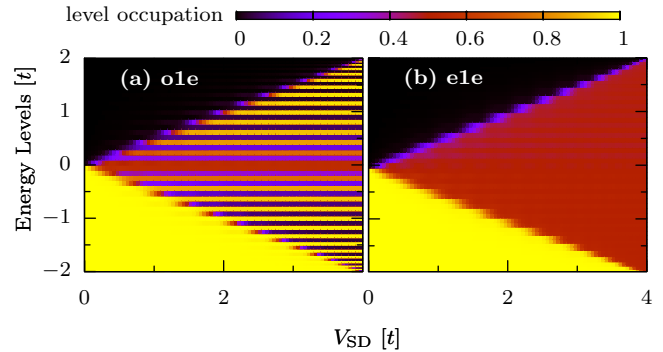}\end{footnotesize}
  \caption{Initial occupation of the single particle energy levels in the non-interacting RLM 
  ($t_\R C=0.4t$) at half filling. The number of lattice sites is $M=M_\R L+M_\R R+1$ with the 
  number of lattice sites in the left (right) lead $M_\R L$ ($M_\R R$). 
  (a) $M_\R L+1=M_\R R=30$. The alternating occupation can be traced back to the alternating 
  localization of the single particle wave functions in either the left or the right lead. 
  (b) $M_\R L = M_\R R=30$. In the uncoupled case ($t_\R C=0$), the energy levels of the leads 
  are degenerate. Therefore the energy levels can not be associated with only one lead.
}
  \label{Fig:RLM_initial_occupation}
\end{center}
\end{figure}

  In a system with an odd number of lattice sites $M$ and spinless electrons, half filling can 
  not be realized strictly since $N=M/2$ is not an integer. Adding spin shifts the particle 
  number at half filling to $N=M$ but leaves a total spin $S^z_\R{tot}=\pm 1/2$, which will 
  occupy the highest single particle level. Since for the doubly occupied levels the spin adds 
  up to 0, the level at the Fermi edge determines the spin density profile which then explains 
  the density peak on the dot in the e1e case and the absence of a peak in the o1o case.
  The time dependent behavior of the current can now be traced back to the single particle 
  energy levels being confined in a single lead (fully delocalized) in the case of 
  different numbers of lattice sites $M_\R L=M_\R R \pm 1$ (equal number of lattice sites 
  $M_\R L = M_\R R$). For the e\textit{n}o and o\textit{n}e configurations, applying a bias 
  voltage as in Eq.~\eqref{eqn:Hamiltonian-SD} leads to an alternating occupation of the 
  energy levels corresponding to the alternating confinement of the single particle wave 
  functions in the left or the right lead. In contrast we find an occupation number of $1/2$ 
  in the energy range $-\VSD/2\ldots\VSD/2$ when $M_\R L=M_\R R$, corresponding to the fully 
  delocalized single particle wave functions. We demonstrate this behavior for the 
  non-interacting resonant level model (RLM) in Fig.~\ref{Fig:RLM_initial_occupation}. 

\begin{figure}[t!]
\begin{center}
  \setlength{\graphiclength}{0.475\textwidth}
  \graphicspath{{./fig_data/data_local_FF_CURRENT/niRLM_Vg/}}
  \begin{footnotesize}\input{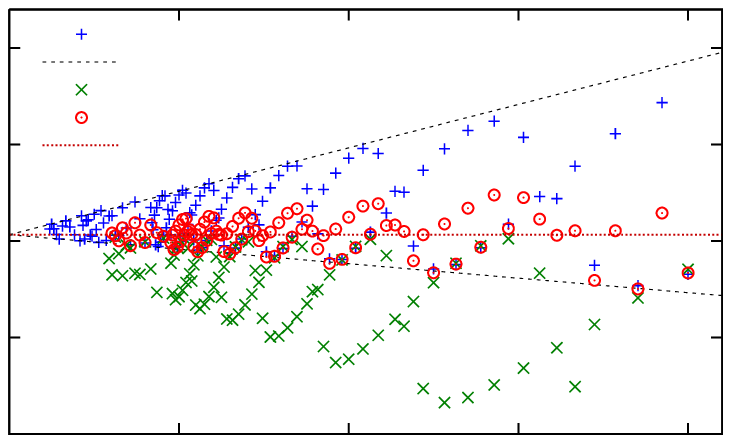}\end{footnotesize}
  \caption{Current through a single impurity with an applied gate voltage $\Vg=0.21t$ for 
  $\VSD=0.5t$, coupled to two leads ($t_\R C=0.3t$), as a function of the system size. The 
  analytic result is obtained using the Landauer--B\"uttiker formula. While for different 
  fillings ($N=M/2$ and $N=M/2-1$) there is a systematic deviation from the analytic result, 
  the interpolation results in a substantial improvement. The linear envelope is plotted to 
  highlight the $1/M$-dependency of the finite size effects. For an explanation of the 
  sinusoidal oscillations see also Fig.~\ref{Fig:RLM_fit_Vg_VSD-dependence} and the text.
}
  \label{Fig:RLM_fit_Vg_M-dependence}
\end{center}
\end{figure}

  So far, we have a connection of the degeneracy of the single particle energy levels for the 
  situation where the impurity is decoupled from the leads with the respective class of the 
  system (e\textit{n}o / o\textit{n}e, o\textit{n}o, e\textit{n}e). The situation changes when 
  adding a constant local potential 
  $\Delta \op V = \Delta V_\R L \op N_\R L + \Delta V_\R R \op N_\R R$ to both, the initial 
  and the time evolution Hamiltonian. 
  To obtain the data of the dotted lines in Fig.~\ref{Fig:twoimpurities_compare_even_odd} we 
  calculated the single particle energy levels for a system with an even (odd) number of 
  lattice sites in the leads and then applied a relative shift of the lead levels with 
  $\Delta V_\R L=-\Delta V_\R R \in \lbrace \epsilon/4, \epsilon/2 \rbrace$ for the two-dot 
  structure and $\Delta V_\R L \in \lbrace \pm \epsilon/2 \rbrace$, $\Delta V_\R R=0$ for the 
  single dot structure, where $\epsilon$ is the energy gap to the first unoccupied energy 
  level. This allows to change the level structure of a certain lead configuration in a way 
  that it resembles one of the other configurations in the vicinity of the Fermi edge without 
  changing the number of lattice sites in the leads. In 
  Fig.~\ref{Fig:twoimpurities_compare_even_odd} we see that the time dependent behavior of the 
  system on the time scale $T<T_\R R$ is only given by the structure of the single particle 
  energy levels that contribute to the current, and the bias voltage $\VSD$, at least as long 
  as we do not include interaction. We therefore conclude that o\textit{n}o as well as 
  e\textit{n}e configurations also can be used to study the I--V-characteristics in the low 
  voltage regime. This may be interesting when investigating structures with an even number of 
  lattice sites on the structure, when the constraint $N=M/2$ has to be fulfilled strictly.
\subsection{Density shift in the leads resulting from finite system size}
\label{Sec:DensityShift}
\begin{figure}[t]
\begin{center}
  \setlength{\graphiclength}{0.475\textwidth}
  \graphicspath{{./fig_data/data_local_FF_CURRENT/niRLM_Vg/}}
  \begin{footnotesize}\input{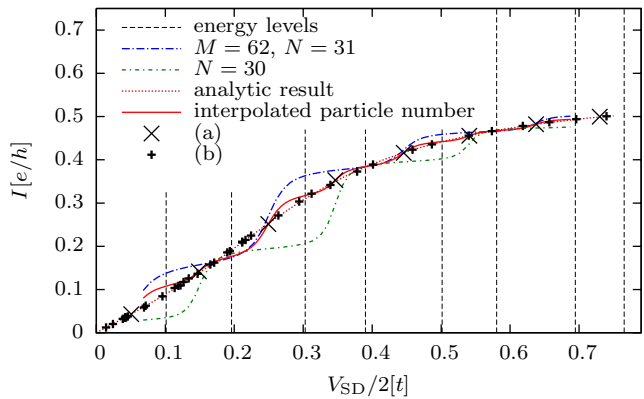}\end{footnotesize}
  \caption{Current through a single impurity with an applied gate voltage $\Vg=0.21t$, coupled 
  to two leads ($t_\R C=0.3t$), as a function of the voltage $\VSD$. The analytic result is 
  obtained using the Landauer--B\"uttiker formula. The vertical lines represent the single 
  particle energies of a system with uncoupled leads ($t_\R C=0.0$); we find that the 
  interpolated value of the current fits best with the analytical result if the bias voltage 
  is chosen as the mean value of two neighboring energy levels (a). However, this condition 
  restricts the bias voltage to only a few values. The restriction can be circumvented by 
  either increasing the number of lattice sites $M$ or by using damped boundary conditions. 
  The latter was used to obtain the values (b) without changing $M$ -- see section 
  \ref{Sec:CorrectSPEL_DBC} for discussion.
}
  \label{Fig:RLM_fit_Vg_VSD-dependence}
\end{center}
\end{figure}

  For the single resonant level model (RLM) the condition of half filling is easily fulfilled 
  by setting the particle number $N=M/2$ as long as the dot level resides in the middle of the 
  band. Then the overall particle number density is $n=1/2$ in the equilibrium case. This can 
  change for different reasons: for example, for a model with two lattice sites in the 
  structure and an overall odd number of lattice sites as discussed before half filling is 
  not realisable, since $M/2$ is not an integer. But even for the RLM, applying a gate 
  voltage $\Vg\neq 0$ changes the particle number on the structure by $\Delta N_\R{Dot}$ while 
  changing the particle number per site in the leads by $-\Delta N_\R{Dot}/(M-1)$ which shifts 
  the lead filling away from $1/2$ as long as the system size $M$ is finite. In this section 
  we will concentrate on the latter case. 

  The impact on the current can be quite large, compare 
  Figs.~\ref{Fig:RLM_fit_Vg_M-dependence}, \ref{Fig:RLM_fit_Vg_VSD-dependence}. 
  The total number of particles must therefore be corrected in such a way that 
  $N_{\text{Leads}}/(M-1)=1/2$ where $N_{\text{Leads}}=N-N_\R{Dot}$ 
  is the particle number in the leads. Thus an initial state $\ket{\Psi_\text i}$ has to be a 
  mixture of states with different particle numbers $\ket{\Psi_{N}}$ and $\ket{\Psi_{N+1}}$, 
  or $\ket{\Psi_{N-1}}$, respectively, depending on the sign of $\Delta N_\R{Dot}$
  \begin{equation}
    \ket{\Psi_\text i}=\alpha \ket{\Psi_{N}} + \beta \ket{\Psi_{N\pm 1}},
  \end{equation} 
  so that 
  \begin{equation}
    \bra{\Psi_\text i} \op N_{\text{Leads}} \ket{\Psi_\text i} = \frac{M-1}2.
  \end{equation} 
  For particle number conserving operators $\op O$ the expectation value reads
  \begin{equation}
    \bra{\Psi_{\text i}}\op O\ket{\Psi_{\text i}}
      = \vert\alpha\vert^2\bra{\Psi_N}\op O\ket{\Psi_N}
       +\vert\beta\vert^2\bra{\Psi_{N\pm1}}\op O\ket{\Psi_{N\pm 1}}
  \end{equation} 
  which leads to the condition
  \begin{eqnarray}
    { \vert\alpha\vert^2\bra{\Psi_N}\op N_{\text{Leads}} \ket{\Psi_N} + ~~~~~~~~~~~~} 
      \nonumber \\ 
    + \vert\beta\vert^2\bra{\Psi_{N\pm1}}\op N_{\text{Leads}}\ket{\Psi_{N\pm 1}} 
    &=&  \frac{M-1}2, \\
    \vert\alpha\vert^2+\vert\beta\vert^2 &=& 1.
  \end{eqnarray}
  Since the current operator $\op I_j$ also is particle number conserving, the resulting time 
  dependent current expectation value is an interpolation of the results for $N$ and for 
  $N\pm 1$ particles in the system
  \begin{equation}
    I_j(T)=\vert\alpha\vert^2 I_j(T;N)+(1-\vert\alpha\vert^2) I_j(T;N\pm 1).
\label{eq:interpolated_current}
  \end{equation} 

  In Fig.~\ref{Fig:RLM_fit_Vg_M-dependence} we show the dependency of the current through a single
  impurity coupled to two leads to the 
  system size for different fillings $N=M/2$ as well as $N=M/2-1$, for a constant value of the 
  bias voltage $\VSD$ and the gate voltage $\Vg$. Furthermore we include the interpolated 
  value, following the procedure described before. We find that the interpolated results are centered around the 
  analytic value, in contrast to the case with fixed particle number. However a 
  distribution with an amplitude $\propto 1/M$ remains. A potential relation of the sinusoidal oscillations in the original 
  data  to the relative position of $\VSD/2$ to the single particle energy 
  levels is illustrated in Fig.~\ref{Fig:RLM_fit_Vg_VSD-dependence}. Here, we show the 
  current as a function of $\VSD$ with $\Vg\neq 0$, where we also apply the interpolation 
  procedure. We compare the analytical result obtained using the Landauer--B\"uttiker approach
  with numerical data for the current through a single impurity coupled to two leads with a
  system size of $M=62$ lattice sites in total. In order to interpolate the current as described
  before, Eq.~\eqref{eq:interpolated_current}, we simulated the time evolution of the current
  expectation value with $N=30$ and $N=31$ particles in the system.
  In comparison to Fig.~\ref{Fig:RLM_fit_Vg_M-dependence} we conclude that one has 
  to choose the system size in relation to the bias voltage carefully to get the desired 
  relation of $\VSD$ and the single particle levels. More precisely, the data points (a),
  that fit nicely with the analytic curve, correspond
  to the interpolated current obtained for a bias voltage where $\VSD/2$ has been chosen as the
  mean value of two neighboring energy levels of the uncoupled ($t_\R C=0$) system.
  Another possibility is the use of damped 
  boundary conditions to shift the single particle levels, which yields the data points (b).
  This idea will be discussed in Section \ref{Sec:CorrectSPEL_DBC}.

  A generalisation of this concept to systems with structures of $M_\R{Dot}>1$ sites with a 
  corresponding number of energy levels is straightforward. A varying gate voltage will 
  change the occupation of the structure in a range $N_\R{Dot}\in[0,M_\R{Dot}]$ with a 
  corresponding change of the particle number in the leads. To get reliable results for the 
  current at half filling in the leads it is then necessary to perform an interpolation of 
  currents with appropriate particle numbers. Results for the linear conductance of a 7-site 
  structure are discussed in the next section.

\section{Results for the conductance} 

\begin{figure}[b]
\begin{center}
\setlength{\graphiclength}{0.475\textwidth}
\includegraphics[width=\graphiclength,clip]{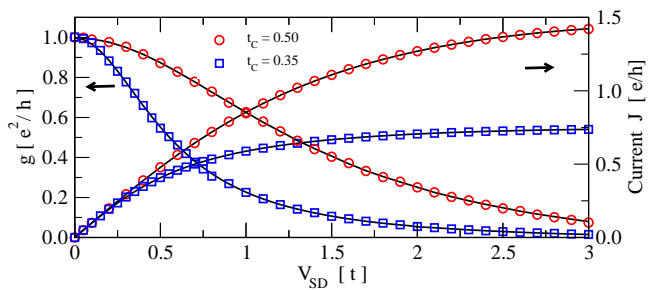}
\caption{
Current and differential conductance as function of applied potential
through a single impurity
with $V_\R g=0$ and half filled leads: $N/M = 0.5$. 
Circles (squares) show results for $t_{\mathrm{C}}=0.5t$ ($0.35t$). 
System size was $M=48$ ($M=96$) and $N_{\mathrm{cut}}=200$ (400)
states were kept in the DMRG.
Lines are exact diagonalization results for $M=512$.
}\label{Fig:Jg-SingleImpurity}
\end{center}
\end{figure}

Our result for the conductance through a single impurity in
Fig.~\ref{Fig:Jg-SingleImpurity} is in excellent quantitative
agreement with exact diagonalization results already for moderate
system sizes and DMRG cutoffs. Accurate calculations for extended
systems with interactions are more difficult, mainly for two reasons: 
1.)~The numerical effort required for our approach depends crucially on
the time to reach a quasi-stationary state.
For the single impurity, the quasi-stationary state is reached on a
timescale proportional to the inverse of the width of the conductance
resonance, $4t\hbar/t_{\mathrm{C}}^2$, in agreement with the result in
Ref.~\cite{wingreen93}.
In general, extended structures with interactions will take longer to
reach a quasi-stationary state, and the time evolution has to be
carried out to correspondingly longer times. 
2.)~In the adaptive td-DMRG, the truncation error grows exponentially
due to the continued application
of the wave function projection, and causes the sudden onset of
an exponentially growing error in the calculated time evolution after
some time. This 'runaway' time is strongly dependent on the DMRG
cutoff, and was first observed in an adaptive td-DMRG study of spin
transport by Gobert et~al.\cite{gobert05}.
To avoid these problems we resort to the full td-DMRG
\cite{peter04}, which does not suffer from the runaway error.

\begin{figure}[t]
\begin{center}
\setlength{\graphiclength}{0.475\textwidth}
\includegraphics[width=\graphiclength,clip]{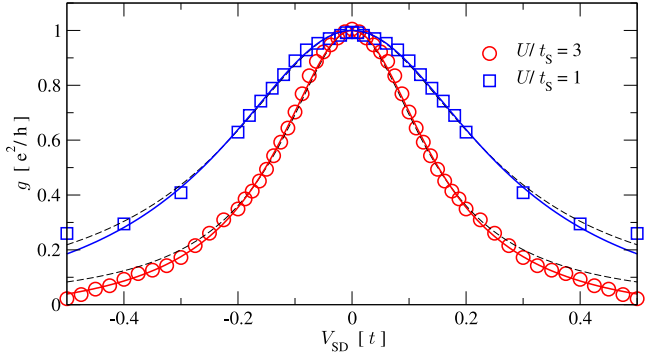}
\caption{
Differential conductance as a function of bias voltage
through a 7 site nanostructure with nearest neighbor interaction. 
Parameters are $t_{\mathrm{C}}=0.5t$, $t_{\mathrm{S}}=0.8t$, and N/M=0.5. 
Squares (circles) denote weak (strong) interaction with
$U/t_{\mathrm{S}}=1\;(3)$ (here: $U_\R C=0.0$). Lines are fits to a Lorentzian with an energy
dependent self energy $\Sigma = {i}\eta_0 + {i}\eta_1 \mu^2$.
Dashed lines: $\eta_1=0$.
System size is $M=144$ ($M=192$) and 600 (800) states were kept
in the DMRG.
}\label{Fig:Jg-d7vX}
\end{center}
\end{figure}

In Fig.~\ref{Fig:Jg-d7vX} we show results for the first differential conductance peak
of an interacting 7-site nanostructure. Careful analysis of the data shows,
that in order to reproduce the  line shape accurately, one has to introduce an
energy dependent self energy for $U/t_{\mathrm{S}}=3$. Since the effect is small,
we approximate it by a correction quadratic in the bias voltage
difference $\mu=\VSD-V_\mathrm{peak}$.
It is important to note that for the strongly interacting nanostructure, $U/t_{\mathrm{S}}=3$,
the conductance peaks are very well separated. Therefore the line
shape {does not overlap with} the neighboring peaks, and the fit is very
robust. Performing the same analysis for a non-interacting nanostructure
with a comparable resonance width, we obtain negligible
corrections to $\eta_1$ in the self energy,
indicating that the change of the line shape is due to correlation effects.

\begin{figure}[t]
\begin{center}
\setlength{\graphiclength}{0.475\textwidth}
\includegraphics[width=\graphiclength,clip]{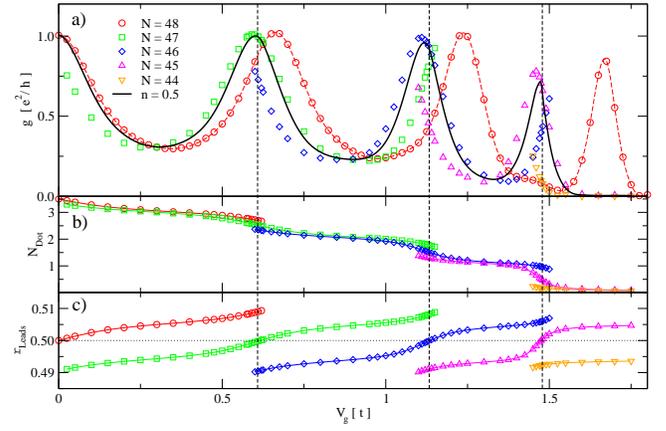}
\caption{
Transport through a non-interacting 7-site nanostructure with
$t_{\mathrm{C}}=0.5t$ and $t_{\mathrm{S}}=0.8t$. 
The energy levels of the nanostructure are indicated by dashed vertical lines.
(a) Linear conductance for different $N$.
The result after applying finite size corrections is shown as 
straight line (see text for details).
(b) Number of fermions on the 7-site nanostructure.
(c) Density $\rho=(N-N_{\mathrm{Dot}})/(M-M_\R{Dot})$ in the leads.
System size is $M=96$ and the number of states kept in the DMRG is
$N_{\mathrm{cut}}=400$.
}\label{Fig:gnf-d7v0t80t50}
\end{center}
\end{figure}

\begin{figure}[t]
\begin{center}
\setlength{\graphiclength}{0.475\textwidth}
\includegraphics[width=\graphiclength,clip]{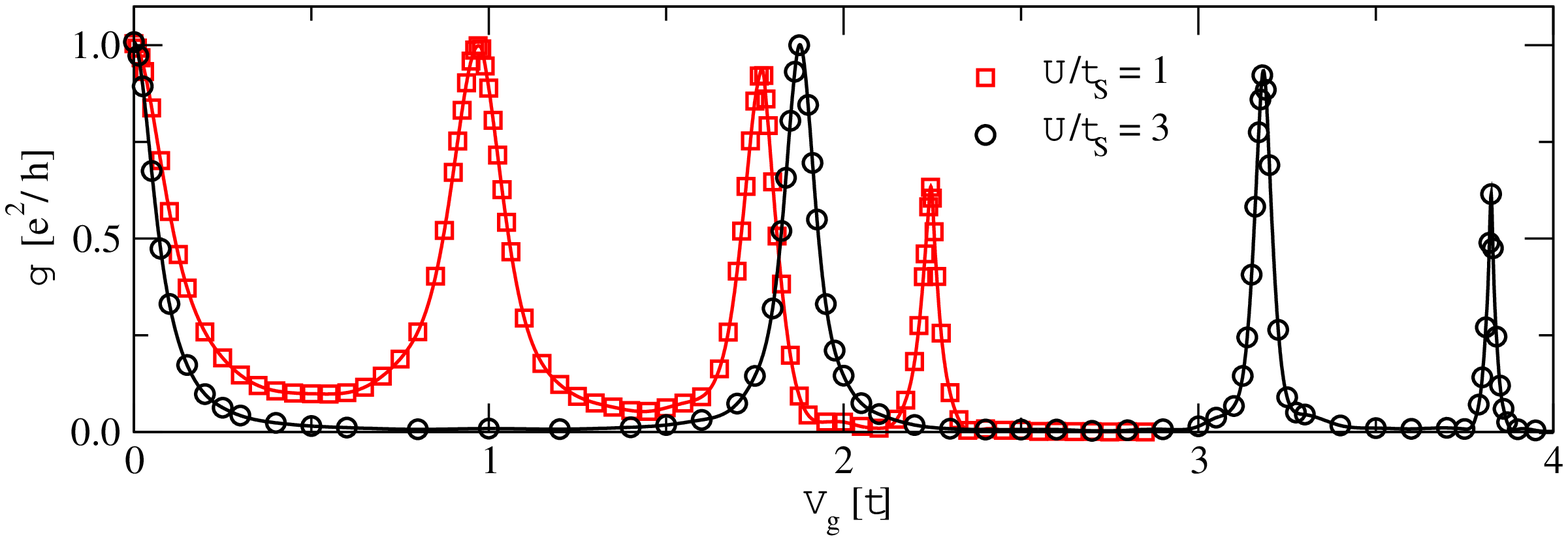}
\caption{
Linear conductance through an interacting 7 site system with
$t_{\mathrm{C}}=0.5t$ and $t_{\mathrm{S}}=0.8t$ for 
weak (squares) and
strong (circles) interaction. System size is $M=96$ ($M=192$) and
400 (600) states were kept in the DMRG.
Finite size corrections have been included.
Lines are guides to the eye.
}\label{Fig:gnf-d7vInt}
\end{center}
\end{figure}

The linear conductance as a function of applied gate potential
can be calculated in the same manner, if
a sufficiently small external potential is used. 
We study the same 7-site nanostructure as before, with interaction $U=0$, and
use a bias voltage of $\VSD=2\cdot10^{-4}$.
For half filled leads, the result for the linear conductance
calculated with a fixed number of fermions, $N/M=0.5$,
is qualitatively correct, but the conductance peaks are shifted to
higher energies relative to the expected peak positions 
at the energy levels of the non-interacting system
(Fig.~\ref{Fig:gnf-d7v0t80t50}). 
Varying the gate potential $\Vg$ increases the charge on the nanostructure
by unity whenever an energy level of the nanostructure moves through the
Fermi level [Fig.~\ref{Fig:gnf-d7v0t80t50} (b)]. The density in
the leads varies accordingly 
[Fig.~\ref{Fig:gnf-d7v0t80t50} (c)]. Since the number
of fermions in the system is restricted to integer values, direct
calculation of the linear conductance at constant $\rho$ is not possible
and one must resort to interpolation.
Using linear interpolation in $\rho(N,\Vg)$ for 
$N=44\dots48$ yields our final result for the linear conductance at
half filling [Fig.~\ref{Fig:gnf-d7v0t80t50} (a)]. The agreement in the
peak positions is well within the expected accuracy for a 96 site
calculation. 
Our results for the conductance through an interacting extended
nanostructure are presented in Fig.~\ref{Fig:gnf-d7vInt}. 
The calculation for the weakly interacting system requires roughly the same
numerical effort as the non-interacting system.
In the strongly interacting case, where the nanostructure is now in the
charge density wave regime, the time to reach a quasi-stationary state
is longer, and a correspondingly larger system size was used in the
calculation.
In both cases we obtain peak heights
for the central and first conductance resonance to within 1\% of the
conductance for a single channel.

\section{Exponential damping}
\label{Sec:ExpDamp}
	In this section we want to study the effect and possible applications of damped boundary 
	conditions (DBC). 
	DBC have been introduced \cite{Vekic_White:PRB1993,dan06} in order to reduce finite size 
	effects.
	Here we would like to reduce the limitations rising from the finite transit time $T_\R R$ 
	and the Josephson wiggling which especially in the low voltage regime and with an applied 
	gate voltage spoils the accuracy of current measurements. 
	We have already seen how to profit from the voltage dependency of the finite size wiggling 
	by using a fit procedure which allows for the calculation of current--voltage 
	characteristics even with an applied gate voltage. We now want to discuss the possibility
	of combining the fit procedure with DBC, where the damping effectively increases the system 
	size. Furthermore we want to use DBC to adjust the single particle energy levels in order 
	to increase the resolution with respect to $\VSD$ when $\Vg\neq 0$, cf. 
	Fig.~\ref{Fig:RLM_fit_Vg_VSD-dependence}.

\begin{figure}[tb]
\begin{center}
  \setlength{\graphiclength}{0.475\textwidth}
  \includegraphics[width=\graphiclength,clip]{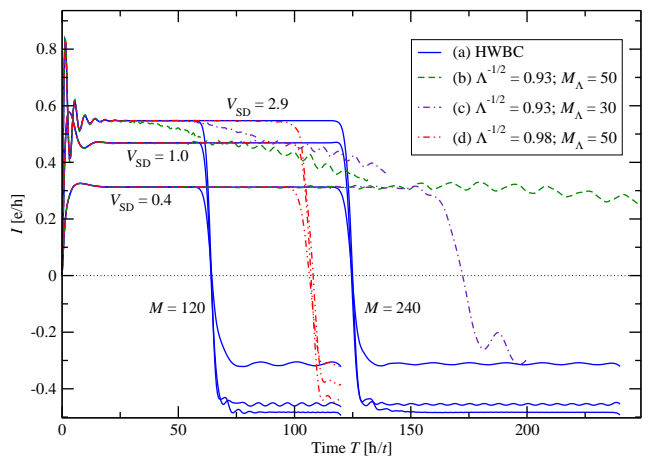}
  \caption{
  Time dependent current through a single impurity with $t_{\mathrm{C}}=0.3t$ at
  nominal half filling $N/M = 0.5$ obtained from exact numerical
  diagonalization for different bias voltages $\VSD$ and different damping conditions. For 
  small bias voltage, finite size reflections from hard wall boundary conditions (HWBC, a) can 
  be suppressed significantly using damped boundary conditions (DBC). Using an exponential 
  damping with $\Lambda^{-1/2}=0.93$, $M=120$ and $M_\Lambda=50$ (b) yields a plateau of 
  constant current for $\VSD=0.4 t$ considerably bigger than in the undamped case. However, 
  the current plateau starts dropping before the estimated transit time according to Eq. 
  \eqref{eq:T_R_in_damped_leads} is reached (here: $T_\R R\approx 670$), which gets even more 
  pronounced when increasing the bias voltage. Reducing the damping (c, d) can lead to good 
  agreement with the estimate ($T_\R R(\R c)\approx 178$, $T_\R R(\R d)\approx 123$).
  }
  \label{Fig:compare_damping_t_0.3}
\end{center}
\end{figure}


\subsection{Estimate for Transit Time in a system with half filling}

\begin{figure}[t]
  \graphicspath{{./fig_data/data_local_FF_CURRENT/test_M_eff/}}
  \begin{footnotesize}\input{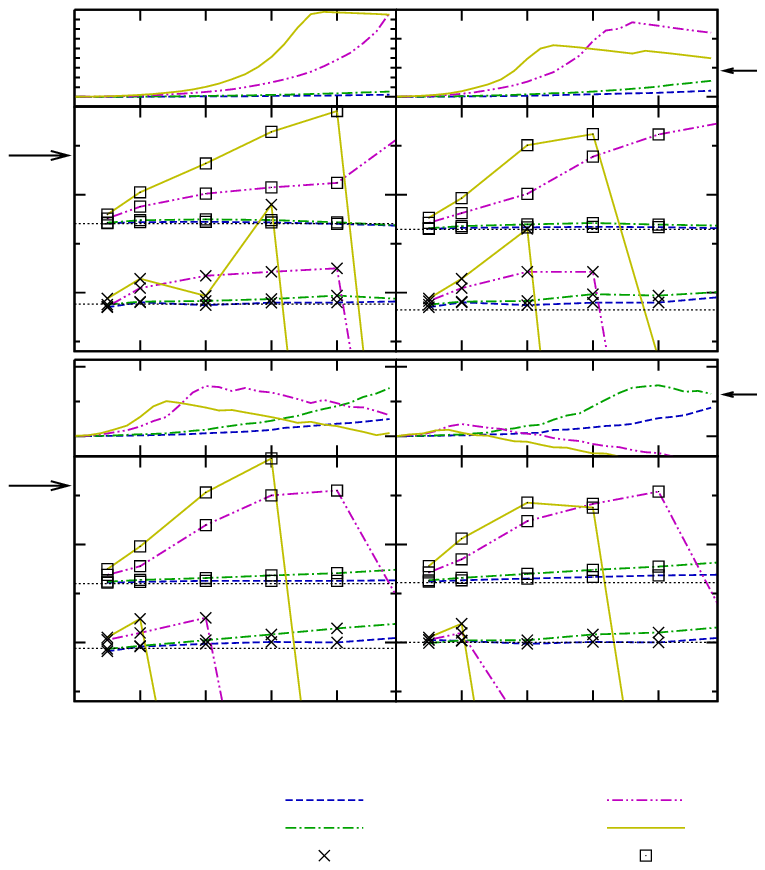}\end{footnotesize}
  \caption
	{
		Test for the transit time estimate $T_\R R$ of the current through a single impurity 
		at half filling, Eqns.~(\ref{eq:T_R},  \ref{eq:T_R_in_damped_leads}), where the black 
		dotted line is the undamped case. All values are plotted as functions of the damped lead 
		size $M_\Lambda$. The small plots at the top show the single particle level density for 
		the energy given by the bias voltage, in units of the level density for the undamped 
		case. (See text for details)
	}
	\label{Fig:T_R_compare}
\end{figure}

\begin{figure}[t]
  \graphicspath{{./fig_data/data_local_FF_CURRENT/Eigenvalues/}}
  \begin{footnotesize}\input{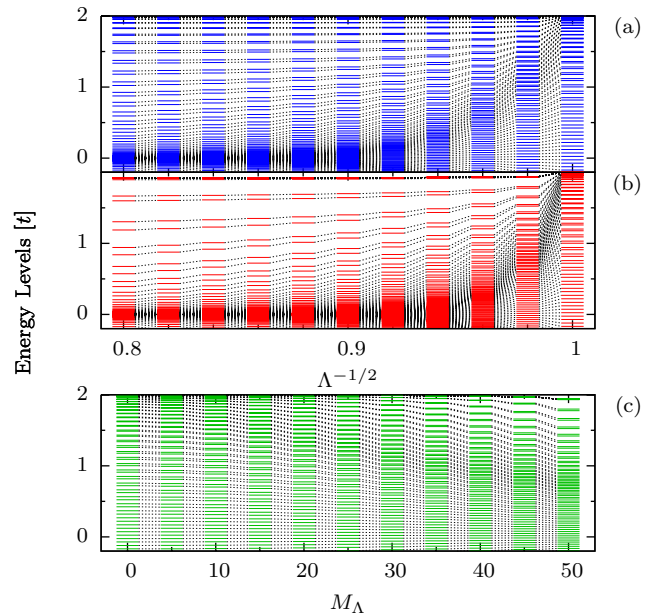}\end{footnotesize}
  \caption{
  Level discretisation in a finite system ($M=120$) with a single impurity, coupled to leads 
  ($t_{\R C}=0.3$) as function of the damping rate $\Lambda^{-1/2}$ (a, b), as well as 
  function of the size $M_\Lambda$ of the damped leads (c). The damping lead size is set to 
  (a) $M_\Lambda=30$ and 
  (b) $M_\Lambda=50$, while for 
  (c) the damping rate is set to $\Lambda^{-1/2}=0.98$. The implementation of damped leads in 
  combination with leads described by a uniform tight binding chain can be used to increase 
  the level density in the vicinity of the fermi edge while allowing for direct access to real 
  space quantities like the current at a specific lattice site, as e.g., the impurity. 
  However, this approach is only useful for the calculation of current in a limited voltage 
  window, since in the high voltage regime also energy levels at the band edge get occupied,
  where the level spacing is significantly increased with $\Lambda$ and $M_\Lambda$.
  }
  \label{Fig:Levelspacing_Lambda_t_0.3}
\end{figure}

  In Fig.~\ref{Fig:compare_damping_t_0.3} we show the time dependent current through a single 
  impurity with $\Vg=0$, including the initial transient regime as well as the finite size 
  reflections for different values of the bias voltage $\VSD$. We compare two different
  system sizes with $M=120$ and $M=240$ lattice sites, and also apply exponentially DBC in order 
  to demonstrate the effectively increased system size. The hopping matrix element is damped towards 
  the boundaries of the system using a damping constant $\Lambda$ as sketched in Fig.~\ref{Fig:damping}, 
  over a range of $M_\Lambda$ lattice sites. 
  The total number of lattice sites is left unchanged (here: $M=120$). 
  We find an enhanced size of the current plateaus,
  however, the damping can also lead to an early breakdown of the current.

  As an estimate for the transit time of a wave packet 
  traveling in undamped leads of size $M$ one can use the Fermi velocity $v_\R F=2t/\hbar$ 
  which leads to
  \begin{equation}
    T_\R R\approx \frac M {v_\R F} = \frac {M\hbar} {2t}.
  \label{eq:T_R}
  \end{equation} 
  Assuming a local Fermi velocity $v_\R F(x)=2t(x)/\hbar$ in damped leads with damping 
  $\Lambda>1$ leads to an expression of the form
  \begin{equation}
    T_\R R\approx \frac {M\hbar} {2t} \left({1-\frac{2M_\Lambda}M}\right) 
      +\frac{2\hbar}{t\ln\Lambda}\big(\Lambda^{M_\Lambda/2}-1\big)
  \label{eq:T_R_in_damped_leads}
  \end{equation} 
  where $M_\Lambda$ is the size of the damped leads. Eq. \eqref{eq:T_R_in_damped_leads} can 
  then be used to estimate an effective system size
  \begin{equation}
    M_{\R{eff}}\approx M-2M_\Lambda+\frac 4{\ln\Lambda}\big(\Lambda^{M_\Lambda/2}-1\big).
  \label{eq:M_eff_in_damped_leads}
  \end{equation} 
  This is in agreement with the results for the pseudo-steady current found for the 
  noninteracting  case, Fig.~\ref{Fig:compare_damping_t_0.3}. For a more quantitative check of 
  the formula we compare the transit time, extracted from a current measurement, to the 
  estimate given by Eq.~\eqref{eq:T_R_in_damped_leads} [Fig.~\ref{Fig:T_R_compare}]. We 
  therefore use two different criteria: (a) the time $T_\R R^\R {(a)}$ where $\dot I(T)$ 
  becomes negative at the end of the first plateau (crosses), and (b) the time 
  $T_\R R^\R {(b)}$ where the current changes sign after one round trip (squares). The black 
  dotted lines show $T_\R R^\R {(a)}$ and $T_\R R^\R {(b)}$ for the undamped case. For values 
  of $\Lambda^{-1/2}$ close to $1$ we find that the estimate is well fulfilled over a wide 
  range of values of $M_\Lambda$ for both (a) and (b) even for big bias voltage. The slight 
  growth of $T_\R R^{(\R a,\R b)}/T_\R R$ is assumed to be caused by the different Fermi 
  velocity of excitations for $\vert \VSD\vert>0$. However, the estimate tends to be totally 
  wrong even for small bias voltage and small values of $M_\Lambda$ if $\Lambda^{-1/2}$ 
  becomes too small.
  The small plots at the top show the relative single particle level density. As expected, cf. 
  Fig.~\ref{Fig:Levelspacing_Lambda_t_0.3}, the level density grows with $M_\Lambda$ until a 
  maximum is reached where the position of the maximum is determined by the bias voltage. 
  It can clearly be seen that the position of the maximum in combination with the values of 
  $T_\R R^{(\R a)}/T_\R R$ gives a strong indication if a current plateau is still well 
  defined for a time scale given by the estimate of $T_\R R$, since 
  $T_\R R^{(\R a)}/T_\R R\simeq 1$ for values of $M_\Lambda$ on the left side of the maximum 
  of the single particle level density. In comparison, (b) is a weak criterion since for 
  strong damping the current plateau starts decaying for times much smaller than $T_\R R$, cf. 
  Fig.~\ref{Fig:compare_damping_t_0.3}.
  In Fig.~\ref{Fig:Levelspacing_Lambda_t_0.3}, we show the single particle energy levels of a 
  system with $M=120$ lattice sites with a single impurity, as function of the damping constant
  $\Lambda^{-1/2}$ as well as function of the size of the damped leads $M_\Lambda$. The plot 
  demonstrates the growth of the level density on the scale $\Lambda^{-M_\Lambda/2}$ which 
  in conjunction with Fig.~\ref{Fig:T_R_compare} allows for an estimate of the maximum value of
  $\VSD$ up to which a current plateau can be expected in a system with DBC.

	\begin{figure}[tb]
		\setlength{\graphiclength}{0.475\textwidth}
		\graphicspath{{./fig_data/data_local_FF_CURRENT/M_eff/}}
		\begin{footnotesize}\input{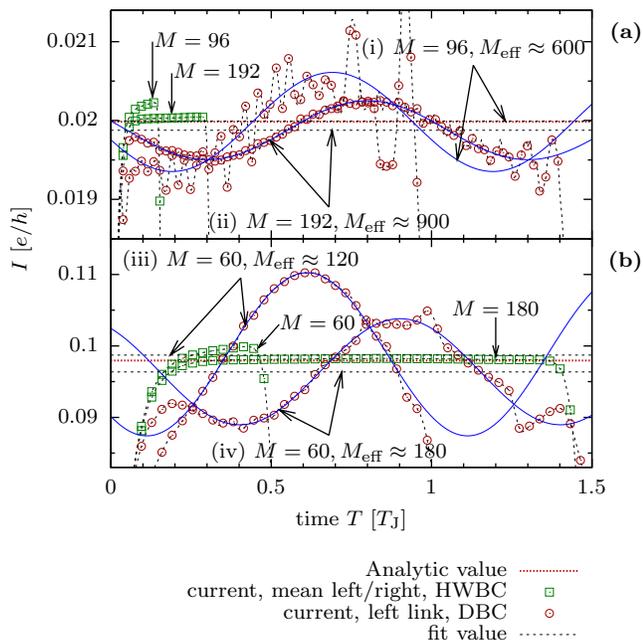}\end{footnotesize}
		\caption{Current through a single impurity with $t_\R C=0.3t$ and $\Vg=0$. The time axis is 
		normalized to the oscillation period $T_\text{J}=2\pi\hbar/\VSD$, with 
		(a) $\VSD=0.02 t$ and 
		(b) $\VSD=0.1t$. The analytic results are computed using the Landauer--B\"uttiker 
		formula. For $\VSD=0.02t$ (a), the oscillation period is $T_\text{J}=314\hbar/t$. To 
		obtain a current plateau containing at least one Josephson oscillation one has to 
		simulate the time evolution of a system with $M\gtrsim 630$, which is very hard on 
		present days computers when interaction is included. Here, we apply DBC on a system with 
		$M=96$ ($M=192$) to effectively increase the system size using 
		(i) $\Lambda\approx 0.903$, $M_\Lambda=32$ 
		(ii, $\Lambda\approx 0.969$, $M_\Lambda=84$). Accidentally, the fit value agrees with 
		the analytic value nearly perfectly for configuration (i). For $\VSD=0.1t$ (b), 
		$T_\text{J}=63\hbar/t \Rightarrow M\gtrsim 126$. The damping conditions are 
		characterized by (iii) $\Lambda\approx 0.93, M_\Lambda=20$ and (iv) 
		$\Lambda\approx 0.900, M_\Lambda=20$, respectively. In both cases one is able to extract 
		a current via the fit procedure although $M \ll 2t T_\text{J}/\hbar$. However, the 
		most reliable results can be obtained by inceasing $M$, c.f. $M=180$ in (b).
		}
	\label{fig:RLM_fit_damping}
	\end{figure}

\subsection{Fit Procedure}

  As already mentioned in Sec.~\ref{sec:finite_size_effects}, the fitting procedure gets 
  unreliable when the oscillation time $T_\text{J}$ substantially exceeds the time range 
  $T_\R S \ldots T_\R R$. We therefore now want to demonstrate how to use the estimate for the 
  transit time in order to implement damping conditions to sufficiently increase the effective 
  system size, enforcing $T_\text{J} \simeq T_\R R-T_\R S$. As an example, we simulate the 
  time evolution of a system with $M$ lattice sites and a single, non-interacting impurity 
  with $V_\R g=0$, and apply a small bias voltage $\VSD>0$. An effective transit time 
  $T_\R R ^\R{eff} \approx T_\R{J}$ can be obtained using DBC, according to 
  Eqns.~(\ref{eq:T_R_in_damped_leads}, \ref{eq:M_eff_in_damped_leads}). 

  The result is presented in Fig.~\ref{fig:RLM_fit_damping}, where we show the time dependent
  current through one of the contact links of a single impurity for different damping conditions
  and two different values of $\VSD$. Again, we fit 
  $\tilde I+\tilde I_\R J \cos(\VSD T+\tilde\varphi)$ to the oscillating part of 
  current expectation value. 
  The extracted current $\tilde I$ for the calculations including DBC fits with the analytic result with 
  an accuracy of $\sim 1 \%$ which is of the same order of magnitude as compared to the mean 
  value of the very small plateau regime that can be found for the system with HWBC. This 
  leads us to the conclusion, that DBC can be used to obtain a first guess while for high 
  precision measurements, HWBC with an increased system size have to be implemented.

\subsection{Correction of the single particle energy levels using DBC}
\label{Sec:CorrectSPEL_DBC}
  In Section \ref{Sec:DensityShift} we found that the effects resulting from a finite density 
  shift in the leads when applying a gate voltage can be significantly suppressed when 
  extracting the current only for certain values of $\VSD$ determined by the single particle 
  level spacing. Since these finite size effects particularly arise in the middle of the band 
  where the density of single particle levels is the lowest -- and where the current has to be 
  extracted for the calculation of the linear conductance -- one would like to shift the 
  levels towards the center of the band somehow. This can be achieved by increasing the number 
  of lattice sites which also increases the numerical effort. 

  Applying DBC also results in a shift of the single particle energy levels in the leads 
  towards the center of the band, cf. Fig.~\ref{Fig:Levelspacing_Lambda_t_0.3}. 
  We therefore state the question if the criterion formulated 
  in Sec. \ref{Sec:DensityShift} still holds true for DBC. The result is shown in 
  Fig.~\ref{Fig:RLM_fit_Vg_VSD-dependence}. To obtain the additional data points (b) we used damping 
  conditions with values of $\Lambda^{-1/2}=0.91\ldots0.98$ and $M_\Lambda=15,20,23$. We calculated
  the single particle energy levels for the decoupled leads and then obtained the current for
  values of the bias voltage with $\VSD/2$ in the middle of two neighboring energy levels.
  To increase the resolution for the high voltage regime only moderate damping conditions are 
  required ($\Lambda^{-1/2}=0.98$, $M_\Lambda=15,20$), while strong damping is imposed to get 
  high resolution in the low voltage regime. For $\VSD$ approaching the band edge, however, 
  DBC have to be avoided for the reasons discussed above.

\section {Conclusions}
We have reviewed the concept of extracting the finite bias and
linear conductance from real time evolution calculations in finite systems.
Very accurate
quantitative results are possible, as long as finite size effects are
taken into account.
Our results for the linear conductance compare favorably both in
accuracy and computational effort with the DMRG evaluation of the Kubo
formula \cite{dan06}.
Calculations of strongly interacting systems show correlation induced
corrections to the resonance line shape.

\begin{acknowledgments} 
We profited from many discussions with Ferdinand Evers, Ralph Werner, and Peter W{\"o}lfle.
We would like to thank Miguel A. Cazalilla for clarifying discussions.
The authors acknowledge the support from the DFG
through project B2.10 of the Center for Functional Nanostructures,
and from the Landesstiftung Baden-W\"{u}rttemberg under project 710.
\end{acknowledgments}

\bibliography{dmrg+1d}


\end{document}